\def\dmix{$D^0-\overline{D}\,^0$ mixing}
\def\d0d0bar{$D^0-\overline{D}\,^0$}
\def\cerenkov{\v Cerenkov\ }
 
\def\bar{\overline}
\def\bra#1{\langle #1 |}

\def\dz{{D^0}}
\def\dzbar{{\overline{D}\,^0}}
\def\dstar{{D^*}}
\def\dstarpl{{D^{*+}}}
\def\dstarmi{{D^{*-}}}
\def\ket#1{| #1 \rangle}

\def\kpl{{K^+}}
\def\kmi{{K^-}}

\def\pipl{{\pi^+}}
\def\pimi{{\pi^-}}

\def\ra{\rightarrow}

\def\genrzdcskbkm{ 0.80_{-1.37}^{+1.46}\pm  0.47}

\def\genrzdcskbkp{ 1.26_{-1.79}^{+1.94}\pm  0.49}

\def\genrzdcskdkm{-0.67_{-1.35}^{+1.44}\pm  0.41}

\def\genrzdcskdkp{ 0.33_{-1.70}^{+1.91}\pm  0.32}

\def\genrzintkbkm{-0.11_{-1.16}^{+1.08}\pm  0.43}

\def\genrzintkbkp{-1.46_{-1.59}^{+1.49}\pm  0.47}

\def\genrzintkdkm{ 0.22_{-1.18}^{+1.12}\pm  0.41}

\def\genrzintkdkp{-0.89_{-1.46}^{+1.35}\pm  0.35}

\def\genrzmixzda{ 0.18_{-0.39}^{+0.43}\pm  0.17}

\def\genrzmixzdabr{ 0.70_{-0.53}^{+0.58}\pm  0.18}

\def\genlimrzmixzda{0.74}
\def\genlimrzmixzdabr{1.45}
\def\gencmixkbpi{0.388}
\def\gencintkbpi{0.499}
\def\gencmixkdpi{0.359}
\def\gencintkdpi{0.473}

\def\likArsddskbkmnos{2269_{ -49}^{+ 49}}

\def\likArsddskbkpnos{2966_{ -56}^{+ 56}}

\def\likArsddskdkmnos{1314_{ -38}^{+ 38}}

\def\likArsddskdkpnos{1677_{ -42}^{+ 42}}

\def\likArsdpikbkmnos{ 746_{ -33}^{+ 33}}

\def\likArsdpikbkpnos{ 797_{ -35}^{+ 35}}

\def\likArsdpikdkmnos{ 311_{ -21}^{+ 21}}

\def\likArsdpikdkpnos{ 368_{ -23}^{+ 23}}

\def\likArsbpikbkmnos{ 338_{ -24}^{+ 24}}

\def\likArsbpikbkpnos{ 423_{ -27}^{+ 27}}

\def\likArsbpikdkmnos{ 278_{ -19}^{+ 19}}

\def\likArsbpikdkpnos{ 356_{ -21}^{+ 21}}

\def\likAwsdpikbkmnos{ 737_{ -32}^{+ 32}}

\def\likAwsdpikbkpnos{ 749_{ -30}^{+ 30}}

\def\likAwsdpikdkmnos{ 323_{ -19}^{+ 19}}

\def\likAwsdpikdkpnos{ 315_{ -19}^{+ 19}}

\def\likAwsbpikbkmnos{ 243_{ -24}^{+ 24}}

\def\likAwsbpikbkpnos{ 333_{ -22}^{+ 22}}

\def\likAwsbpikdkmnos{ 214_{ -16}^{+ 16}}

\def\likAwsbpikdkpnos{ 238_{ -17}^{+ 17}}

\def\likslopezkbkmnos{-3.29_{-0.81}^{+0.81}}

\def\likslopezkbkpnos{-3.73_{-0.68}^{+0.68}}

\def\likslopezkdkmnos{-2.86_{-0.76}^{+0.76}}

\def\likslopezkdkpnos{-2.65_{-0.70}^{+0.70}}

\def\likdazmassnos{1865.8_{-0.1}^{+0.1}}

\def\likKbpizsigmanos{15.16_{-0.16}^{+0.16}}

\def\likKdpizsigmanos{10.76_{-0.15}^{+0.15}}

\def\likdstarzqanos{ 5.92_{-0.01}^{+0.01}}

\def\likdstarzsigmnos{ 0.76_{-0.01}^{+0.01}}
\def\likrzdcskbkm{ 0.90_{-1.09}^{+1.20}\pm  0.44}
\def\likrzdcskbkmnos{ 0.90_{-1.09}^{+1.20}}

\def\likrzdcskdkm{-0.20_{-1.06}^{+1.17}\pm  0.35}
\def\likrzdcskdkmnos{-0.20_{-1.06}^{+1.17}}

\def\likrzintkbkm{-0.46_{-0.97}^{+0.89}\pm  0.41}

\def\likrzintkbkp{-0.84_{-1.00}^{+0.92}\pm  0.43}

\def\likrzintkdkm{-0.29_{-0.96}^{+0.89}\pm  0.37}

\def\likrzintkdkp{-0.25_{-0.94}^{+0.87}\pm  0.36}

\def\likrzmixzda{ 0.39_{-0.32}^{+0.36}\pm  0.16}
\def\likrzmixzdanos{ 0.39_{-0.32}^{+0.36}}

\def\likAzrefzkmnos{  60_{ -14}^{+ 15}}

\def\likAzrefzkpnos{  49_{ -14}^{+ 15}}
\def\liksyskbpref{0.02}
\def\liksyskbpsta{0.27}
\def\liksyskbpbin{0.25}
\def\liksyskbptim{0.17}
\def\liksyskbpmas{0.18}
\def\liksyskbptot{0.44}
\def\liksyskdpref{0.02}
\def\liksyskdpsta{0.17}
\def\liksyskdpbin{0.21}
\def\liksyskdptim{0.19}
\def\liksyskdpmas{0.11}
\def\liksyskdptot{0.35}
\def\liksysmixref{0.01}
\def\liksysmixsta{0.09}
\def\liksysmixbin{0.11}
\def\liksysmixtim{0.06}
\def\liksysmixmas{0.05}
\def\liksysmixtot{0.16}
\def\liklimrzmixzda{0.85}

\def\ncirzmixzda{ 0.21_{-0.09}^{+0.09}\pm  0.02}

\def\nomrzdcskbkm{ 0.68_{-0.33}^{+0.34}\pm  0.07}
\def\nomrzdcskbkmnos{ 0.68_{-0.33}^{+0.34}}

\def\nomrzdcskdkm{ 0.25_{-0.34}^{+0.36}\pm  0.03}
\def\nomrzdcskdkmnos{ 0.25_{-0.34}^{+0.36}}

\def\nomsyskbpref{0.04}
\def\nomsyskbpsta{0.02}
\def\nomsyskbpbin{0.01}
\def\nomsyskbptim{0.01}
\def\nomsyskbpmas{0.04}
\def\nomsyskbptot{0.07}
\def\nomsyskdpref{0.00}
\def\nomsyskdpsta{0.02}
\def\nomsyskdpbin{0.00}
\def\nomsyskdptim{0.01}
\def\nomsyskdpmas{0.02}
\def\nomsyskdptot{0.03}

\def\likcoefdada{ 1.00}

\def\likcoefdcda{ 0.58}
\def\likcoefdcdc{ 1.00}

\def\likcoefdeda{-0.92}
\def\likcoefdedc{-0.68}
\def\likcoefdede{ 1.00}

\def\likcoefdfda{-0.90}
\def\likcoefdfdc{-0.68}
\def\likcoefdfde{ 0.95}
\def\likcoefdfdf{ 1.00}

\def\likcoefdgda{-0.71}
\def\likcoefdgdc{-0.90}
\def\likcoefdgde{ 0.84}
\def\likcoefdgdf{ 0.85}
\def\likcoefdgdg{ 1.00}

\def\likcoefdhda{-0.70}
\def\likcoefdhdc{-0.90}
\def\likcoefdhde{ 0.82}
\def\likcoefdhdf{ 0.83}
\def\likcoefdhdg{ 0.95}
\def\likcoefdhdh{ 1.00}

\def\likcoefdida{ 0.78}
\def\likcoefdidc{ 0.74}
\def\likcoefdide{-0.92}
\def\likcoefdidf{-0.93}
\def\likcoefdidg{-0.92}
\def\likcoefdidh{-0.90}
\def\likcoefdidi{ 1.00}

\documentstyle[preprint,prd,aps]{revtex}

\begin{document}
\draft
\input psfig.sty

\title{A search for \d0d0bar\ mixing \\
and doubly-Cabibbo-suppressed decays of the $D^0$ \\
in hadronic final states}

%
%
\author{
    E.~M.~Aitala,$^9$
       S.~Amato,$^1$
    J.~C.~Anjos,$^1$
    J.~A.~Appel,$^5$
       D.~Ashery,$^{15}$
       S.~Banerjee,$^5$
       I.~Bediaga,$^1$
       G.~Blaylock,$^8$
    S.~B.~Bracker,$^{16}$
    P.~R.~Burchat,$^{14}$
    R.~A.~Burnstein,$^6$
       T.~Carter,$^5$
 H.~S.~Carvalho,$^{1}$
  N.~K.~Copty,$^{13}$
    L.~M.~Cremaldi,$^9$
 C.~Darling,$^{19}$
       A.~Fernandez,$^{12}$
       P.~Gagnon,$^2$
       C.~Gobel,$^1$
       K.~Gounder,$^9$
     A.~M.~Halling,$^5$
       G.~Herrera,$^4$
 G.~Hurvits,$^{15}$
       C.~James,$^5$
    P.~A.~Kasper,$^6$
       S.~Kwan,$^5$
    D.~C.~Langs,$^{11}$
       J.~Leslie,$^2$
       B.~Lundberg,$^5$
       S.~MayTal-Beck,$^{15}$
       B.~Meadows,$^3$
 J.~R.~T.~de~Mello~Neto,$^1$
    R.~H.~Milburn,$^{17}$
 J.~M.~de~Miranda,$^1$
       A.~Napier,$^{17}$
       A.~Nguyen,$^7$
  A.~B.~d'Oliveira,$^{3,12}$
       K.~O'Shaughnessy,$^2$
    K.~C.~Peng,$^6$
    L.~P.~Perera,$^3$
    M.~V.~Purohit,$^{13}$
       B.~Quinn,$^9$
       S.~Radeztsky,$^{18}$
       A.~Rafatian,$^9$
    N.~W.~Reay,$^7$
    J.~J.~Reidy,$^9$
    A.~C.~dos Reis,$^1$
    H.~A.~Rubin,$^6$
 A.~K.~S.~Santha,$^3$
 A.~F.~S.~Santoro,$^1$
       A.~J.~Schwartz,$^{11}$
       M.~Sheaff,$^{18}$
    R.~A.~Sidwell,$^7$
    A.~J.~Slaughter,$^{19}$
    M.~D.~Sokoloff,$^3$
       N.~R.~Stanton,$^7$
       K.~Stenson,$^{18}$
    D.~J.~Summers,$^9$
 S.~Takach,$^{19}$
       K.~Thorne,$^5$
    A.~K.~Tripathi,$^{10}$
       S.~Watanabe,$^{18}$
 R.~Weiss-Babai,$^{15}$
       J.~Wiener,$^{11}$
       N.~Witchey,$^7$
       E.~Wolin,$^{19}$
       D.~Yi,$^9$
       S. Yoshida,$^{7}$                         
       R.~Zaliznyak,$^{14}$
       and
       C.~Zhang$^7$ \\
(Fermilab E791 Collaboration)
}

\address{
$^1$ Centro Brasileiro de Pesquisas F{\'i}sicas, Rio de Janeiro, Brazil\\
$^2$ University of California, Santa Cruz, California 95064\\
$^3$ University of Cincinnati, Cincinnati, Ohio 45221\\
$^4$ CINVESTAV, Mexico\\
$^5$ Fermilab, Batavia, Illinois 60510\\
$^6$ Illinois Institute of Technology, Chicago, Illinois 60616\\
$^7$ Kansas State University, Manhattan, Kansas 66506\\
$^8$ University of Massachusetts, Amherst, Massachusetts 01003\\
$^9$ University of Mississippi, University, Mississippi 38677\\
$^{10}$ The Ohio State University, Columbus, Ohio 43210\\
$^{11}$ Princeton University, Princeton, New Jersey 08544\\
$^{12}$ Universidad Autonoma de Puebla, Mexico\\
$^{13}$ University of South Carolina, Columbia, South Carolina 29208\\
$^{14}$ Stanford University, Stanford, California 94305\\
$^{15}$ Tel Aviv University, Tel Aviv, Israel\\
$^{16}$ 317 Belsize Drive, Toronto, Canada\\
$^{17}$ Tufts University, Medford, Massachusetts 02155\\
$^{18}$ University of Wisconsin, Madison, Wisconsin 53706\\
$^{19}$ Yale University, New Haven, Connecticut 06511\\
}

\date{\today}
\maketitle

\begin{abstract}
	We present results of a search for \dmix\ 
        and doubly-Cabibbo-suppressed decays of the $D^0$ 
	in Fermilab experiment E791, a fixed-target charm hadroproduction
	experiment. We look for evidence of mixing in the decay
	chain $\dstar\ra\pi D\ra\pi(K\pi$ or $K\pi\pi\pi)$. If the 
	charge of the pion from the $\dstar$ decay 
	is the same as the charge of the kaon
	from the $D$ decay (a ``wrong-sign'' event), 
	mixing may have occurred. Mixing can be distinguished
	from other sources of wrong-sign events (such as
	doubly-Cabibbo-suppressed decays) by analyzing
	the distribution of decay times. We see no evidence
	of  mixing. Allowing for CP violation in the interference 
        between DCS and mixing amplitudes our fitted
        ratio for mixed to unmixed decay rates is
	$r_{mix}=(\likrzmixzda)\%$.
        This corresponds to a 90\%\ CL upper limit of 
	$r_{mix} < \liklimrzmixzda\%$.
	The sensitivity of this result is comparable to that
	of previous measurements, but the assumptions made in
	fitting the data are notably more general. We present 
	results from many fits to our data under various assumptions. 
        If we assume $r_{mix} = 0$, we find a two-sigma wrong-sign enhancement
        in the $K\pi$ mode which we ascribe to doubly 
        Cabibbo-suppressed decays. The ratios of doubly
        Cabibbo-suppressed decays to Cabibbo-favored decays are 
        $r_{dcs}(K\pi)      =(\nomrzdcskbkm)\%$ and
        $r_{dcs}(K\pi\pi\pi)=(\nomrzdcskdkm)\%$.
\end{abstract}

\section{Introduction}

The Standard Model predicts a rate for \dmix\ 
which
is many orders of magnitude below the reach of present experiments.
Typical calculations \cite{mixpred} give $r_{mix}$, the ratio of
mixed to unmixed decay rates, in the range
$10^{-10}$\ to 10$^{-7}$.
In contrast, various extensions to the Standard Model \cite{beyond}
allow a mixing 
rate close to the current experimental sensitivity
of $10^{-3}$ to $10^{-2}$.
Consequently, a discovery of \dmix\ at currently
measurable levels would be 
inconsistent with the Standard Model, and 
would provide a clear signal for new physics.

Experimentally, mixing is identified by a change
in the charm quantum number 
of the neutral $D$ meson between its production and decay.
In the analysis presented in this paper,
the charm of
the produced $D$ is determined from the decay
$\dstarpl\ra\dz\pipl$ (or $\dstarmi\ra\dzbar\pimi$),
where the charge of the pion indicates whether
a $\dz$ or a $\dzbar$ was produced.
$D$ decays are reconstructed
in four all-charged
hadronic decay modes 
$D\ra \kmi\pipl$, 
$D\ra \kpl\pimi$,
$D\ra \kmi\pimi\pipl\pipl$ or 
$D\ra \kpl\pipl\pimi\pimi$. 
(Hereafter, we will omit the charge superscripts
from the final states where context allows.)
Possible evidence for mixing comes from the detection of a 
meson produced as 
a $\dz$ ($\dzbar$) decaying to a ``wrong-sign'' final state 
which contains a
$\kpl$ ($\kmi$), with the kaon charge opposite to that expected for
unmixed decays. 

Fermilab experiment E691 \cite{E691}
has previously used this technique to set 
what is currently the strictest upper limit on mixing, 
$r_{mix}<0.37\%$, albeit with specific assumptions which we will
address in this paper. Fermilab experiment E615 obtained
a limit of $r_{mix}<0.56\%$ by looking for same-sign muon pairs
in $\pi$-tungsten interactions, 
based on a specific model for charm production
\cite{E615}. 
Evidence for wrong-sign decays has been presented by 
the CLEO collaboration \cite{CLEO}, which measures
the ratio of wrong-sign to right-sign decays
to be ($0.77\pm 0.25\pm 0.25$)\%  for the $K\pi$ final state.
However, the CLEO 
experiment was unable to distinguish between mixing and
doubly-Cabibbo-suppressed decays, which also produce
wrong-sign events. Recently we have reported on a search for mixing
using semileptonic decays of the $\dz$ ($\dzbar$) which do not have a
doubly-Cabibbo-suppressed background\cite{semi791}. 
We found that $r_{mix}<0.50\%$.

It is possible to distinguish doubly-Cabibbo-suppressed (DCS) 
and mixing contributions
to the wrong-sign rate by studying the distribution of $D$ decay times.
In the limit of small mixing,
the rate for wrong-sign decays takes the form
	\begin{eqnarray}
	\Gamma[\dz(t)\ra f]&=&{e^{-\Gamma t}\over4}
	|\bra{f}H\ket{\dzbar}_{CF}|^2
	\left|{q\over p}\right|^2 \nonumber \\[3pt]
	\times \Bigl[4|\lambda|^2 &+&\left((\Delta M)^2+
        {(\Delta\Gamma)^2\over4}\right)t^2+\bigl(2Re(\lambda)\Delta\Gamma
	+4Im(\lambda)\Delta M\bigr)t\Bigr],
	\label{ratef}
	\end{eqnarray}
where
	\begin{equation}
	\lambda\equiv\ {p\over q}
	{\bra{f}H\ket{\dz}_{DCS}\over \bra{f}H\ket{\dzbar}_{CF}},
	\label{deflambda}
	\end{equation}
and $p$ and $q$ 
describe the relationship between the 
charm eigenstates $\ket{\dz}$ and $\ket{\dzbar}$ and
the mass eigenstates $\ket{D_{1,2}}$:
	\begin{eqnarray}
	|D_1\rangle&=& p|\dz\rangle+q|\dzbar\rangle,\nonumber\\
	|D_2\rangle&=& p|\dz\rangle-q|\dzbar\rangle.
	\label{defpq}
	\end{eqnarray}
The amplitude $\bra{f}H\ket{\dz}_{DCS}$ represents
the DCS decay of the $\dz$ while $\bra{f}H\ket{\dzbar}_{CF}$ is the
Cabibbo-favored counterpart for the decay of $\dzbar$.
The parameters $\Delta M$ and $\Delta\Gamma$ describe the differences
in mass and width of the two physical states. 
The term proportional to 
$|\lambda|^2$ in Eq. (\ref{ratef}) describes the contribution from
DCS amplitudes, the term proportional to $t^2$ describes the
lowest-order contribution from mixing, and the term proportional to
$t$ represents the interference between mixing and DCS amplitudes.
We can apply this formula to the measured time distribution
of wrong-sign decays to determine the separate
contributions from
DCS and mixing amplitudes.

In the study that follows, we examine a sample of about 9,100
reconstructed, tagged $\dz$ decays to look for wrong-sign decays, using
the different time distributions to separate the DCS and mixing
contributions in our search. As we shall see, there are no significant
wrong-sign signals in our data, which leads us to set restrictions on
the ratio of wrong-sign to right-sign rates. The most likely fit (in the
possible Standard Model scenarios) will be presented first.  Afterwards,
we will determine the effects of relaxing all constraints and of
additional constraints (absence of DCS--mixing interference, no mixing
at all) which investigate interesting physics cases or are necessary to
compare with previously published results.

\section{Effects of CP Violation}\label{sect_CP}
 
Equation~(\ref{ratef}) describes the rate for $\dz$ to decay to a
wrong-sign final state $f$. Within the context of some new physics
models, it is possible that the rate for $\dzbar$ to decay to $\bar f$
is not the same, and that CP is violated to a significant extent.
Thus, it is important to allow for the possibility of CP violation. This
results in the most conservative upper limit on wrong-sign decays. The
analysis presented here is the first experimental study to allow for
the possibility of CP violation.
(For recent discussions of the role of CP violation in \dmix\ see
\cite{Blaylock}, \cite{Wolfenstein}.)

Formally, the conjugate equation is
	\begin{eqnarray}
	\Gamma[\dzbar(t)\ra\bar f]&=&{e^{-\Gamma t}\over4}
	|\bra{\bar f}H\ket{\dz}_{CF}|^2
	\left|{p\over q}\right|^2 \nonumber \\[3pt]
	\times \Bigl[4|\bar\lambda|^2 &+&\left((\Delta M)^2+
	{(\Delta\Gamma)^2\over4}\right)t^2+\bigl(2Re(\bar\lambda)\Delta\Gamma
	+4Im(\bar\lambda)\Delta M\bigr)t\Bigr],
	\label{ratefbar}
	\end{eqnarray}
with
	\begin{equation}
	\bar\lambda\equiv\ {q\over p}
	{\bra{\bar f}H\ket{\dzbar}_{DCS}\over \bra{\bar f}H\ket{\dz}_{CF}}.
	\label{deflambdabar}
	\end{equation}
In principle, CP violation can arise through a difference 
between Equations~(\ref{ratef}) and (\ref{ratefbar}) in any one of the
three terms. Any term in (\ref{ratef}) 
can differ from its charge conjugate in (\ref{ratefbar}) as a 
result of the interference of two or more contributing amplitudes
which have non-zero relative phases of both the CP-conserving
and CP-violating type.

Inequality of the two constant terms
({\it i.e.}, $\left|{q\over p}\right|^2|\lambda|^2\neq
\left|{p\over q}\right|^2|\bar\lambda|^2$, but see comment \cite{CFCPinv}) 
is referred to as $direct$ CP violation. This could be 
significant if two or more comparable DCS amplitudes contribute
with different CP-conserving and CP-violating phases.
However, the Standard Model contribution (which is expected to
dominate) provides only one weak, CP-violating phase.
Direct CP violation is therefore likely to be small.
Similarly, the two charge conjugate terms proportional to $t^2$
will be the same unless there are two or more mixing
amplitudes with relative CP-violating and CP-conserving phases. 
On the contrary,
most models suggest that if mixing occurs at all,
it is likely to be dominated by a single
CP-violating phase. 
Therefore, the most plausible constraint involving CP violation
restricts CP violation to the interference term.
We will explore this possibility, as well as the more general
case without CP restrictions, 
in the study of our data which follows.
 
\section{Description of E791}
 
      We report the results of a search for \dmix\ and DCS decays 
using hadronic decays found in data from our experiment, Fermilab 
E791. We collected approximately $2\times 10^{10}$ hadronic interactions in the
1991-2 fixed-target run using the TPL spectrometer \cite{TPL}
with a 500 GeV/$c$ $\pi^-$ beam.
There were five foil targets: one $0.5$-mm 
thick Pt foil followed by four $1.6$-mm thick diamond foils with
15 mm center-to-center separations. This arrangement allowed us to
greatly reduce secondary interaction backgrounds by selecting only charm
candidates which decayed in air.
 
      The target region was preceded by six planes of silicon 
microstrip detectors and 8 proportional wire chambers (PWC's) 
used for beam tracking and was followed by 17 additional planes of
silicon microstrip detectors for measuring tracks produced at 
and downstream of the
primary vertex. The track momenta and slopes were also measured in
the downstream spectrometer which had two magnets, 
35 planes of drift chambers, and two PWC's.
Two threshold \cerenkov counters provided
$\pi$/$K$ separation in the 6 - 60 GeV/$c$ momentum
range \cite{Bartlett}.
 
      The mixing analysis in this paper
relies heavily on track reconstruction, which
begins by using hits in the silicon detector and folds in additional
information from the downstream devices. The tracking efficiency is
approximately 80\%\ for particles with momenta greater than 30 GeV/$c$ 
and drops to
around 60\%\ for particle momenta of 10 GeV/$c$.
The mean number of reconstructed tracks used to fit the primary vertex
is seven.
After reconstruction, events with evidence of multiple vertices were
kept for further analysis.
The list of reconstructed vertices is used in the selection criteria
described below. 

      We determined our production (primary) and decay vertex
resolutions by comparing reconstructed and true vertex positions using
our Monte Carlo detector simulation. 
The transverse resolutions quoted below are
one-dimensional values.  Longitudinal and transverse position
resolutions for the primary vertex are 350 and 6 $\mu m$, respectively.
For the mean $D^0$ momentum of 65 GeV/$c$ the longitudinal resolutions for
$K\pi$ and $K\pi\pi\pi$ vertices are 320 and 395 $\mu$m, respectively, and
increase by 33 and 36 $\mu$m, respectively, for every 10 GeV/$c$ $D^0$
momentum. 
Similarly, for the mean momentum, 65 GeV/$c$, of the observed $D^0$'s, 
the transverse
resolutions for $K\pi$ and $K\pi\pi\pi$ vertices are 10 and 12 $\mu$m 
respectively and decrease by about 0.5 $\mu$m for every 10 GeV/$c$
increase in $D^0$ momentum.
 
The kaon and pion
identification efficiencies and misidentification 
probabilities vary with momentum
and with the signatures we require in the \cerenkov detector.
For typical particle momenta in the
range 20 GeV/$c$ to 40 GeV/$c$, the
\cerenkov identification efficiency of a kaon is around 58\%\ when the
probability for a pion to be misidentified as a kaon is 4\%.
In the same momentum range the
\cerenkov identification efficiency of a pion is around 93\%\ when the
probability for a kaon to be misidentified as a pion is 35\%.

\section{Selection of Data Sample}\label{sect_sample}
 
A search for the rare wrong-sign mixing and
DCS decays requires 
selection
criteria that
emphasize background reduction. 
We achieve this goal in two stages: initially reconstructing
displaced secondary vertices and using a few 
loose criteria for selecting $\dz$ decays to reduce the data sample,
and then optimizing the data selection with artificial neural networks.
 
Initial reduction of the large E791 data set 
to a manageable size was achieved with the aid of a
few simple criteria.
Here, we describe the cuts made in
these initial stages for the $D^0 \to K\pi$ mode.
(When we refer to $K\pi$ or $K\pi\pi\pi$ in this paper, without any
explicit signs, we include the charge conjugate states. Otherwise, we
indicate a specific final state by explicitly specifying at least the
kaon charge or specifying right-sign (RS) or wrong-sign (WS) decays.)
Two-prong vertices 
were used to start the search for $\dz$ decays.
The invariant mass of the two-prong
$\dz$ candidate, assumed to be $K\pi$,
was required to be in the range 1.7 to 2.0
GeV/$c^2$. The kaon candidate was chosen as the one with the higher
probability of being a kaon based on 
\cerenkov information.
To further reduce the contributions from misidentified
$\dz$ decays, $K\pi$ candidates were rejected if the reverse
hypothesis ($\pi K$) fell within 2 $\sigma$ of the $\dz$ mass, where
$\sigma$ is the measurement resolution for the $D^0$ mass.
Similarly, to reduce contamination from $\dz$ decays to $K^+K^-$ and
$\pi^+\pi^-$, $K\pi$ candidates were rejected if the $K^+K^-$ or
$\pi^+\pi^-$ mass hypotheses fell within 2 $\sigma$ of the $\dz$ mass.
To help ensure that the reconstructed secondary vertex was a true
decay vertex, we required that it be separated from the
primary vertex by at least 8 standard deviations ($\sigma_{\Delta z}$) 
in the beam direction
({\it i.e.}, $\Delta z/\sigma_{\Delta z} > 8$). 
Two further requirements ensured that the reconstructed $\dz$
was consistent with originating at the primary vertex.
First, the impact parameter of the reconstructed $D$ momentum
with respect to the primary vertex, $b_p$, was required to 
be less than 60 $\mu$m. Second, the component of
the reconstructed $D$ momentum perpendicular to the $D$ line-of-flight
(as determined from the primary and secondary vertex positions),
$p_T^D$,
was required to be less than 0.35 GeV/$c$.
The $K$ and $\pi$ decay tracks were
required to be well-reconstructed in the silicon detectors and drift chambers.
Finally, 
the momentum asymmetry of the $K$ and $\pi$ as measured in the
laboratory frame ({\it i.e.}, 
$p_{asy}\equiv |\vec p_K - \vec p_\pi|/|\vec p_K + \vec p_\pi|$)
was required to be
less than 0.65. This reduced the contribution from random track
combinations, which tended to be asymmetric. 

The cuts for the
$D^0\to K\pi\pi\pi$ case were similar in
the initial stages. 
We used candidates arising from both 
4-prong vertices 
and 3-prong vertices (with an added track). 
The two vertex samples contributed roughly equal
amounts to the signal. 
We required $\Delta z/\sigma_{\Delta z}>8$, $b_p< 60$~$\mu$m,
$p_T^D< 0.5$ GeV/$c$, and that the decay vertex was outside the target foils.
The $D^0$ candidate mass was required to be in the range 1.7 - 2.0 GeV/$c^2$.
To eliminate reflections from Cabibbo-favored decays 
we examined the hypothesis that the kaon was actually a pion and one of
the pions opposite in charge to the kaon was actually a kaon. Since
there are two such possibilities, candidate $K\pi\pi\pi$ decays were
rejected if either possibility yielded a candidate mass within 2$\sigma$
of the $\dz$ mass.
Tracks were required to have momenta greater than 0.5 GeV/$c$ 
and to have greater than $4\sigma$ of transverse separation 
from the primary vertex where $\sigma$ is the measurement
resolution on the separation. 
Finally, the decay tracks were
required to form a vertex 
that was
no more than 2.5 cm downstream of the final target.
Beyond that point the silicon detectors and other material in 
the beam path provided large numbers of secondary interactions.
 
      In the final stage of analysis, we used two-layer feed-forward neural
networks to optimize the signal selection \cite{Beale and Jackson,Haykin}.
Specifically, we chose selection criteria that maximize $S/\sqrt{B}$ 
where $S$ and $B$ were the signal 
and the background under the signal for the right-sign decays.
A vector, whose components are variables such as the ones just mentioned,
was fed into each neural net as the input layer. Each node in the next (hidden)
layer computed the sigmoid of the sum of an offset and the inner product
of the input vector with a weight vector. The results from this layer in
turn formed the input for a single node in the final (output) layer.
Thus, the networks effectively combined information from each variable we would
otherwise have ``cut'' on and provided a single output value in the range
0 to 1. This output was monotonically
related to the probability that a given candidate was signal and not
background.
 
Since our two major sources of
background were 
false $D^0$
candidates and real $D^0$ candidates combined with random pions to
produce fake $D^*$ candidates, we used separate neural networks to
classify the $D^0$ and $D^*$ candidates. Although there are only two
modes of $D^0$ decay examined in this work, three
$D^0$ samples were used
to train the neural nets: one for the $D^0\to K\pi$ mode and two for
the $D^0\to K\pi\pi\pi$ mode.
The two separate $D^0\to K\pi\pi\pi$ samples contained 
candidates from vertices that had either all four or only 
three of the four tracks. 
 
      In order to minimize our dependence on Monte Carlo, we used $D^0$
candidates in our real data to train separate neural nets for each of
the three samples. We chose $D^0$ candidates that do not combine with
pions to give a $D^*$ candidate. The training sample is thus independent
of our mixing sample and ensures that the neural net training was
unbiased. A fourth neural net was trained
using part of the right-sign $D^{*+} \to D^0 \pi^+$ sample to classify $D^{*+}$
candidates. Every net was trained using events in the peak region as
``signal'' and the remaining events as ``background''. 
We selected only those events for which the product of
the $D^0$ and $D^*$ net outputs was greater than a certain value rather
than making individual ``cuts'' on many variables.
 
      In the $D^0\to K^-\pi^+$ mode, the net was presented with twelve
input variables:
the $p_T$ of the $D^0$ relative to the incident pion beam direction,
the separation between the secondary and primary vertices ($\Delta z$),
$p_T^D$,
$b_p$,
the $\chi^2$ per degree of freedom for the secondary vertex fit,
the \cerenkov-based probability for the kaon to be a kaon,
the momentum asymmetry ($p_{asy}$),
the consistency probability for the secondary vertex to be in a target foil,
the track fit $\chi^2$ per degree of freedom for the two tracks 
and the number of tracking systems 
traversed by each of
the two tracks. The two nets for the $D^0\to K^-\pi^+\pi^+\pi^-$ mode
used seven variables, of which the first six variables were the same as
the first six just listed for the $D^0\to K^-\pi^+$ mode.
The seventh variable was the 
smallest contribution to the $\chi^2$ of the fit to the primary vertex
from any of the four decay tracks.
Finally, the $D^*$ neural
net was constructed using only five variables: the $\chi^2$ per degree
of freedom 
for the track fit for the pion
from the $D^{*+}$ decay (referred to hereafter as the ``bachelor
pion''), the number of tracking stations in which the pion is detected,
the probability that it is not a fictitious track,
its momentum and the $\chi^2$ per degree of freedom
for the vertex fit of the $D^0$ and the bachelor pion.
There were three nodes in the hidden layer for the $D^0 \to K^-\pi^+$
and $D^*$ nets and four nodes in the hidden layers of each of the
$D^0\to K^-\pi^+\pi^+\pi^-$ nets.
 
      Although we considered many variables, we pruned 
the list down to the variables listed above 
and also pruned some of
the connections to the hidden layer nodes when their contributions to the
output were deemed unimportant using a
technique called subset reduction, implemented as follows.
Nodes in a given
layer were viewed as a linear array, one row for each event. The matrix
thus formed was subjected to
singular value decomposition using QRcp factorization\cite{Strang}.
The ``energy content'' of the nodes was determined by the
resulting
eigenvalues \cite{Kanjilal}. Nodes with an ``energy content'' $<$ 1\%\
were deleted. 
 
      We also tried other techniques for selecting events, including the
more common method of using independent ``cuts'' in each variable and 
a Binary Decision Tree (BDT) technique \cite{BDT}.
The sensitivity of the neural net technique was about 10\%\
higher than the BDT in the $D^0 \to K\pi$ mode and about 30\%\
higher than the BDT in the $D^0\to K\pi\pi\pi$ mode; in turn the
BDT was better than the commonly used ``cuts'' technique.
One further advantage of the neural net technique was that the
output could be used to choose the best candidate in an event, should
there be more than one (a rare occurence). This simplified the
statistical analysis in the fits to our data.
 
\begin{figure}
	\centering
	\centerline{\psfig{file=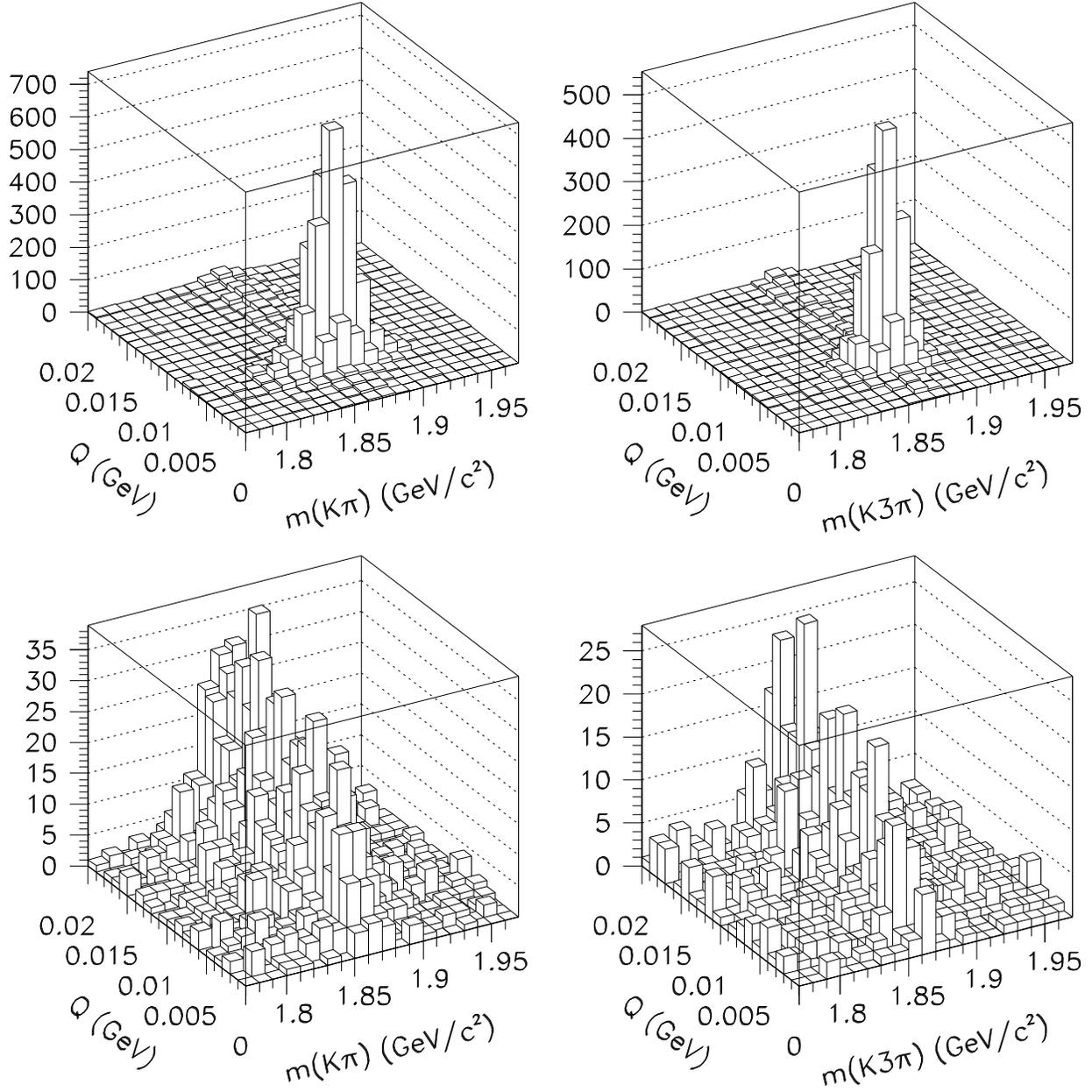,height=17cm,angle=0}}
	\vskip 1.0cm
	\caption{Plots of $Q$ (defined in the text) versus
	the candidate $D$ mass for right-sign $D\ra K\pi$
	(top-left), right-sign $D\ra K\pi\pi\pi$ (top-right),
	wrong-sign $D\ra K\pi$ (bottom-left), and 
	wrong-sign $D\ra K\pi\pi\pi$ (bottom-right). Clean signals
	are apparent in both right-sign plots. In all four plots,
	the bands of events at $m(K\pi),\ m(K\pi\pi\pi)\approx 1.87$
	GeV/$c^2$ are due to real $D$ decays combining with random
	pions to give false $\dstar$ candidates.}
        \label{fglego}
\end{figure}

The results of our neural net optimizations are shown in Figure~\ref{fglego}
for right-sign and wrong-sign $K\pi$ and $K\pi\pi\pi$ final states.
In this figure, we plot the candidate $\dz$ mass ($m(K\pi)$ or $m(K\pi\pi\pi$))
versus $Q$, defined as $Q\equiv m(K\pi\pi)- m(K\pi)-m(\pi)$ 
or $Q\equiv m(K\pi\pi\pi\pi) - m(K\pi\pi\pi) - m(\pi)$.
For real $\dstar$ decays, $Q$ has a value of about 5.8 MeV.
In the right-sign plots (top of Figure~\ref{fglego}), clear signals
are apparent over small backgrounds. The bands of events at
$m(K\pi)$ or $m(K\pi\pi\pi)\approx 1.865$ GeV/$c^2$ are due to real $\dz$
decays combining with random pions in the event
to give a false $\dstar$ candidate. These bands are more readily
seen in the wrong-sign plots (bottom of Figure~\ref{fglego})
where the vertical scale is expanded by a factor of 20.
This background, which we 
will refer to as ``random pion'' background, is the dominant one 
in our analysis. 
We will call the remaining broad background visible in the
plots the ``false $D^0$'' background.

In the right-sign plot for the $K\pi$ mode, 
there is a signal of 5643 events above a
background of 235 events.
In the right-sign plot for the $K\pi\pi\pi$ mode, 
there is a signal of 3469 events above a background of 146 events. 
The signals and backgrounds are estimated in a region 
spanning $\pm 1.75\sigma$ around the peak in $Q$,
for $1.77 < m_{D^0} < 1.97$ GeV/$c^2$. 
The precise region used to estimate $S$ and $B$ is not important for the 
optimization. The resulting sensitivities ($S/ \sqrt{B}$) 
for the two modes are 368 ($K\pi$) and 287 ($K\pi\pi\pi$).

\section{Fit Technique}\label{sect_technique}
 
The process just described results in eight separate datasets: $\dz$ or
$\dzbar$, decaying to $K\pi$ or $K\pi\pi\pi$, right-sign or wrong-sign
decays. Although it is possible for us to fit each data set separately
(which we have done as a check), it is useful to combine all eight data
sets into a single fit. This allows us to take advantage of the fact
that the central values of the $D$ and $D^*$ signals, as well as the
$m_{K\pi}$, $m_{K\pi\pi\pi}$ and Q resolutions, are the same for the
different data samples. Under these circumstances, most of the
parameters of the single fit (which are largely parameters to describe
background) remain uncoupled, and in that sense are no different from
eight separate fits. Only the signal masses and resolutions are
constrained across data samples. Studies of separate fits for the
different samples show no significant shifts from the single fit results
and have convinced us that these constraints are valid.

Our most general fit includes no constraints beyond those just
described, and is summarized in Section \ref{sect_other} 
and in Table \ref{tbfitgen}.  
However, as discussed in Section \ref{sect_CP}, the most likely scenario
is that there is no CP violation in either the DCS or the pure
mixing terms of the wrong-sign rates. This leads to three additional
constraints, discussed at the end of this section, which then lead to
the results of Table \ref{tbfitlik}. These results are the main focus
of our studies in mixing. Finally, in Section \ref{sect_other} we perform 
other fits using additional physical restrictions 
(no DCS--mixing interference, or no mixing at all)
in order to explore other physics
hypotheses and to compare with previous measurements. In what follows,
we describe the terms of the fit in detail.

As stated, we perform a single unbinned maximum likelihood fit to
the data, using the following form for the $\ln$ of the likelihood:
	\begin{equation}
	\ln{{\cal L}} =  \sum_i \ln{{\cal L}_i}
        - \sum_f \left({1\over2}\ln{[2\pi N_{pred}^f]} 
	+ {[N_{pred}^f - N_{obs}^f]^2\over 2 N_{pred}^f}\right),
	\label{eqlike}
	\end{equation}
where the first sum is over all the $\dstar$ candidates,
the second sum is over the eight decays used in the analysis, 
${\cal L}_i$ represents the likelihood for each candidate, 
and $N_{obs}^f$ is the observed number of candidates for each final state.
The argument of the second sum is the
logarithm of a normalized Gaussian, and serves to constrain the
number of candidates predicted by the fit, $N_{pred}^f$.
There are three contributions to each ${\cal L}_i$: signal,
random pions with real $\dz$'s, and random pions with false $\dz$'s.
In addition, for the $\dz\rightarrow K^\mp\pi^\pm$ samples we include a
contribution from misidentified $\dz\rightarrow K^+K^-$ and 
$\dz\rightarrow\pi^+\pi^-$
decays, which also contribute measurable background.
	\begin{equation}
	{\cal L}_i = S(m_i,Q_i,t_i) + M(m_i,Q_i,t_i) 
	+ P(m_i,Q_i,t_i) 
	+ F(m_i,Q_i,t_i),
	\end{equation}
where $m_i$, $Q_i$, and $t_i$ are the $D$ mass, $Q$ value and
proper decay time of each candidate.
A wrong-sign signal event is described by simple Gaussian terms
in $m_i$ and $Q_i$, multiplied by 
a sum of the three different decay time distributions
that represent the DCS, mixing and interference contributions 
(see Eq.~(\ref{ratef})):
	\begin{eqnarray}
	\label{eqwssig}
	S(m_i,Q_i,t_i) &=& {1\over N_{pred}}\ 
	{1\over\sqrt{2\pi}\sigma_D}
	e^{-(m_\dz-m_i)^2 / 2\sigma_D^2}
	\times	
	{1\over\sqrt{2\pi}\sigma_Q}
	e^{-(Q_\dstar-Q_i)^2 / 2\sigma_Q^2}
	\times \nonumber \\
	&&\Bigl\{
	A_{dcs}\ B_{exp}(t_i) + A_{mix}\ B_{mix}(t_i) 
	+ A_{int}\ B_{int}(t_i)
	\Bigr\},
	\end{eqnarray}
where
	\begin{eqnarray}
	\label{Bdef}
        B_{exp}(t_i) &=& B_{exp}^0\ \epsilon(t_i)\
                  \ \int dt\ e^{-(t-t_i)^2/2\sigma_0^2}\
                e^{-\Gamma t},\nonumber\\
        B_{mix}(t_i) &=& B_{mix}^0\ \epsilon(t_i)
                  \ \int dt\ e^{-(t-t_i)^2/2\sigma_0^2}\
                t^2\ e^{-\Gamma t},\\
        B_{int}(t_i) &=& B_{int}^0\ \epsilon(t_i)
                  \ \int dt\ e^{-(t-t_i)^2/2\sigma_0^2}\
                t\ e^{-\Gamma t},\nonumber
	\end{eqnarray}
$\epsilon(t_i)$ is the reconstruction efficiency and $\sigma_0$ is the
decay time resolution. 
Each $B(t_i)$ is normalized to unit integral so that
$A_{dcs}$, $A_{mix}$, and $A_{int}$ can be
interpreted as the number of $observed$ candidates of each type. The
Gaussian smearing integrals are performed analytically with a smearing
width $\sigma_0$ = 0.05 ps.
 
The reconstruction efficiency $\epsilon(t)$ is the first of three
functions that must be modeled for the fit.
It is desirable to measure
this function
using real data rather than using Monte Carlo simulation.
Fortunately, this can be accomplished with
a sample of right-sign events.
Since there is no mixing contribution to
the right-sign decay rate,
the true decay time distribution 
for right-sign decays is proportional to
$e^{-\Gamma t}$ with $\Gamma = (0.415\ {\rm ps})^{-1}$
\cite{PDG}.
The reconstructed distribution is proportional to 
$\epsilon(t)\,\int dt\ e^{-(t-t_i)^2/2\sigma_0^2}\
                e^{-\Gamma t}$.
Therefore, dividing the 
measured right-sign distribution 
(corrected for non-$\dz$ background using sideband subtraction)
by the known smeared exponential
gives a distribution proportional to the efficiency\cite{tsmear}.
Figure~\ref{fgefficiencies} shows the results of that measurement
for both the $K\pi$ and $K\pi\pi\pi$ final states, which we will use 
to represent $\epsilon(t)$ in the fit.

      Despite the explicit mass cuts designed to reduce backgrounds from
$\dz \to KK$ and $\dz \to \pi\pi$ decays described in Section
\ref{sect_sample}, some contamination remains.
The misidentified $\dz \to KK$ and $\dz \to \pi\pi$ events are described by 
	\begin{equation}
        \label{refeqn}
        M(m_i,Q_i,t_i) = {1\over N_{pred}}\ A_{KK,\pi\pi}\
	U(m_i)\
	V(Q_i)\ B_{exp}(t_i).
	\label{eqwsrefl}
	\end{equation}
where the functions $U(m_i)$ and $V(Q_i)$ are obtained from Monte
Carlo simulations of $KK$ and $\pi\pi$ reflections remaining after all
cuts, including explicit mass cuts designed to minimize these
reflections. The parameters $A_{KK,\pi\pi}$ describe the number of
events in the wrong-sign $K^+\pi^-$ and $K^-\pi^+$ samples.
Similar backgrounds for the $K\pi\pi\pi$ mode are not significant (see
Section \ref{sect_syst} for further discussion).

The random pion background is described by
	\begin{equation}
        \label{rpieqn}
	P(m_i,Q_i,t_i) = {1\over N_{pred}}\ A_\pi\
	{1\over\sqrt{2\pi}\,\sigma_D}e^{-(m_\dz-m_i)^2 / 2\sigma_D^2}\
	R(Q_i)\ B_{exp}(t_i).
	\label{eqwsranpi}
	\end{equation}
The background shape in $Q$, represented by $R(Q)$, is 
independent of the candidate $D$ mass.
We model this shape by
combining a $\dz$
candidate from one event with a $\pi$ from another
event. As long as
the $\dz$ is not strongly correlated with other 
tracks in the event, and the selection cuts are not dependent
on the spatial relation between the $\dz$ and other tracks in
the event, this technique should provide an accurate model
of background. Monte Carlo studies confirm the validity of
this method.
The resulting distribution is
compared to the wrong-sign $\dz\ra K\pi$ and $\dz\ra K\pi\pi\pi$ 
data samples in
Figure~\ref{fgR(Q)}.

\begin{figure}
	\centering
	\centerline{\psfig{file=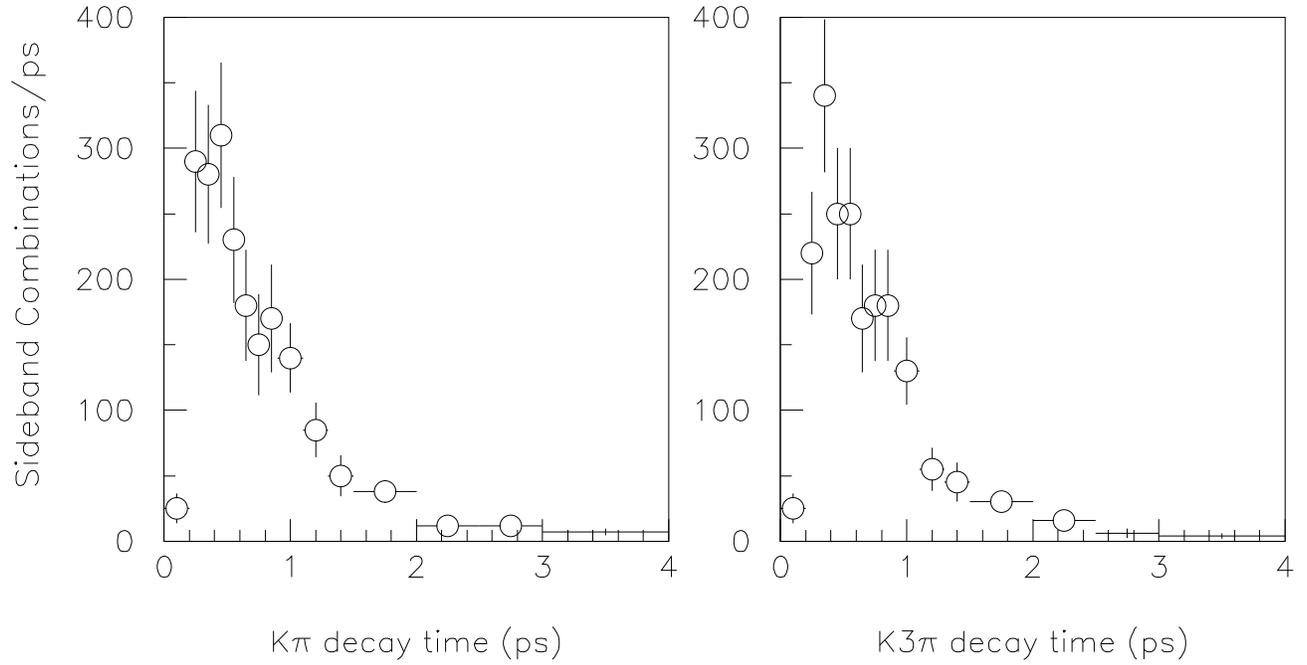,height=10cm,angle=0}}
	\vskip 0.5cm
	\caption{The measured decay time distribution
	for $\dz\ra K\pi$ candidates (left)
	and $\dz\ra K\pi\pi\pi$ candidates (right), taken from the sidebands
	$1.77<m_{D}<1.81$ GeV/$c^2$ and $1.93<m_{D}<1.97$ GeV/$c^2$.
	These distributions are used to represent $B_{false}(t)$
	in the likelihood fit.}
	\label{fgbcomb}
\end{figure}

The false $D^0$ background is adequately described by a linear function
in $m_i$:
	\begin{equation}
	F(m_i,Q_i,t_i) = {1\over N_{pred}}\ 
	{A_0(1 + A_1(m_i-m_0))\over \Delta m}\ 
        R(Q_i)\ B_{false}(t_i),
	\end{equation}
where $m_0$ is an arbitrary reference chosen to be 1.87 GeV/$c^2$ and
$\Delta m$ is the $D^0$ mass interval (0.2 GeV$/c^2$). 
The function $R(Q)$ is
observed
to be the same as in the case of the random pion and real $D^0$
background described by Equation~(\ref{rpieqn}) above.
The function $B_{false}(t_i)$ describes the time distribution of 
the false $D^0$ background. We model this distribution
using candidates from the $D$ mass sidebands of the right-sign event sample
(Figure~\ref{fgbcomb}).
Since this background is very small, we do not need to model it with great
precision, and the statistics of Figure~\ref{fgbcomb} are adequate.
 
The likelihood function for right-sign decays is constructed similarly.
Since right-sign decays were used to model $\epsilon(t)$ and $B_{false}(t)$,
we do not use the lifetime information for these events in the fit. Moreover,
right-sign decays are not subject to mixing or interference, so the fit
functions for 
these events are given by the simplified formulae
	\begin{eqnarray}
	S(m_i,Q_i) &=& {1\over N_{pred}}\ 
	{1\over\sqrt{2\pi}\,\sigma_D}
	e^{-(m_\dz-m_i)^2 / 2\sigma_D^2}
	\times	
	{1\over\sqrt{2\pi}\,\sigma_Q}
	e^{-(Q_\dstar-Q_i)^2 / 2\sigma_Q^2}
	\times A_{rs}, \nonumber \\
	P(m_i,Q_i) &=& {1\over N_{pred}}\ A_\pi\
	{1\over\sqrt{2\pi}\,\sigma_D}e^{-(m_\dz-m_i)^2 / 2\sigma_D^2}\
	R(Q_i), \\
	F(m_i,Q_i) &=& {1\over N_{pred}}\ 
	{A_0(1 + A_1(m_i-m_0))\over \Delta m}\
        R(Q_i). \nonumber
	\end{eqnarray}

We fit all the data, both right-sign and wrong-sign 
$\dz\ra K\pi$ and $\dz\ra K\pi\pi\pi$,
simultaneously in one fit. Separate terms for charge conjugate final
states are provided to allow for the most general possible form for
CP violation. 
Under these conditions, we have four signal parameters
($A_{dcs}$, $A_{mix}$, $A_{int}$ and $A_{KK,\pi\pi}$) and
three background parameters ($A_\pi$, $A_0$ and $A_1$) for the two 
wrong-sign decay modes $\dz\ra\kpl\pimi$ and $\dzbar\ra\kmi\pipl$.
We have three signal parameters
($A_{dcs}$, $A_{mix}$ and $A_{int}$) and
three background parameters ($A_\pi$, $A_0$ and $A_1$) for the two 
wrong-sign decay modes $\dz\ra\kpl \pi\pi\pi$ and $\dzbar\ra\kmi \pi\pi\pi$.
For each right-sign mode 
($\dzbar\ra\kpl\pimi$, $\dzbar\ra\kpl \pi\pi\pi$,
$\dz\ra\kmi\pipl$, and $\dz\ra\kmi \pi\pi\pi$)
we have one signal parameter ($A_{rs}$)
and three background parameters
($A_\pi$, $A_0$ and $A_1$). Additionally, we have five mass
parameters ($m_\dz$, $\sigma_{K\pi}$, $\sigma_{K\pi\pi\pi}$,
$Q_\dstar$, $\sigma_Q$)
to describe the signal Gaussian functions.
Separate $\dz$ mass resolutions are used for the $K\pi$ and $K\pi\pi\pi$
final states. The resolution in $Q$ is dominated by the bachelor pion, and is
therefore the same for the two final states. 
 
With this list we have 47 parameters for a complete description
of the data. However, we expect that the false $D^0$ backgrounds for 
right-sign and wrong-sign decays to the same $\dz$ final state should have
the same slope parameter ($A_1$), although the level ($A_0$) may differ
since they are combined with pions of different charge to form the
$\dstar$ candidates. This observation 
reduces the number of parameters to 43.
 
We also note that the values $r_{mix}(\dz\ra\dzbar)$ and 
$r_{mix}(\dzbar\ra\dz)$ should be independent of the $D$ decay final state.
Thus, without loss of generality, we can assume that 
$r_{mix}(\dz\ra\kpl\pimi) = r_{mix}(\dz\ra\kpl \pi\pi\pi)$ and
$r_{mix}(\dzbar\ra\kmi\pipl) = r_{mix}(\dzbar\ra\kmi \pi\pi\pi)$.
This eliminates two more parameters from our fit \cite{weightedsum},
leaving us
with a total of 41 independent parameters to describe the full data set.

It is convenient to express the wrong-sign signal parameters
$A_{dcs}$, $A_{mix}$ and $A_{int}$ in terms of the ratios of produced
wrong-sign events to produced right-sign events,
since these are the parameters of primary physics interest.
For the wrong-sign $\kmi\pipl$ final state
	\begin{eqnarray}
	\label{eqmixratios}
	r_{dcs}(\kmi\pipl)  &=& {A_{dcs}(\dzbar\ra\kmi\pipl) \over 
		A_{rs}(\dzbar\ra\kpl\pimi)},\nonumber\\
	r_{mix}(\kmi\pipl) &=& {A_{mix}(\dzbar\ra\kmi\pipl)\over 
		A_{rs}(\dzbar\ra\kpl\pimi)}\times c_{mix},\\
	r_{int}(\kmi\pipl) &=& {A_{int}(\dzbar\ra\kmi\pipl)\over 
		A_{rs}(\dzbar\ra\kpl\pimi)}\times c_{int},\nonumber
	\end{eqnarray}
and similarly for $\kpl\pimi$, $\kmi\pi\pi\pi$ and $\kpl\pi\pi\pi$.
The $c$'s in the expressions for $r_{mix}$ and $r_{int}$ are given by
	\begin{eqnarray}
	\label{eqeffcorr}
	c_{mix} &=& {\int dt_i\,\Gamma \epsilon(t_i) 
                     \int dt\,e^{-\Gamma t}\ e^{-(t-t_i)^2/2\sigma_0^2}\ \over 
		\int dt_i\,{1\over2}\Gamma^3\ \epsilon(t_i) 
                \int dt\,t^2\,e^{-\Gamma t}\ e^{-(t-t_i)^2/2\sigma_0^2}},\\
	c_{int} &=& {\int dt_i\,\Gamma \epsilon(t_i) 
                     \int dt\,e^{-\Gamma t}\ e^{-(t-t_i)^2/2\sigma_0^2}\ \over
                \int dt_i\,\Gamma^2\ \epsilon(t_i) 
                \int dt\,t  \,e^{-\Gamma t}\ e^{-(t-t_i)^2/2\sigma_0^2}}.
\nonumber
	\end{eqnarray}
These terms
correct for the different integrated efficiencies for reconstructing 
wrong-sign DCS, mixing and interference
events. Table~\ref{tbeffcorr} shows these correction factors for 
both the $K\pi$ and $K\pi\pi\pi$ final states.
 
Although the production characteristics 
of $\dz$ and $\dzbar$ are different in our
experiment, the ratios in Equation~(\ref{eqmixratios})
are designed to cancel this effect.
In constructing these ratios we implicitly assume that the
Cabibbo-favored amplitudes $\bra{\bar f} H \ket{\dz}$ and 
$\bra{f} H \ket{\dzbar}$ are equal in magnitude, as mentioned previously
\cite{CFCPinv}.
With this assumption, $r_{mix}$ of Equation~(\ref{eqmixratios}) 
can be interpreted according to convention as
	\begin{equation}
	r_{mix} = {1\over 2\Gamma^2}\left|{q\over p}\right|^2
	\left((\Delta M)^2+{(\Delta\Gamma)^2\over4}\right).
	\end{equation}
 
At this point, the fit is completely general, with no physics assumptions
applied. However, as discussed in Section \ref{sect_CP}, it is unlikely
that CP is violated in the DCS and pure mixing terms of the wrong-sign rates,
even in most extensions to the standard model.
Under these circumstances, there are three additional constraints, namely
$r_{dcs}(\dz\ra\kpl\pimi)=r_{dcs}(\dzbar\ra\kmi\pipl)$,
$r_{dcs}(\dz\ra\kpl\pi\pi\pi)=r_{dcs}(\dzbar\ra\kmi\pi\pi\pi)$,
and $r_{mix}(\dz\ra\dzbar)=r_{mix}(\dzbar\ra\dz)$. These constraints
remove three more parameters from the fit, leaving a total of 38.
We will use this fit to give us our primary result, summarized in 
Table \ref{tbfitlik}.
 
\section{Results}
 
\begin{figure}
	\centering
	\centerline{\psfig{file=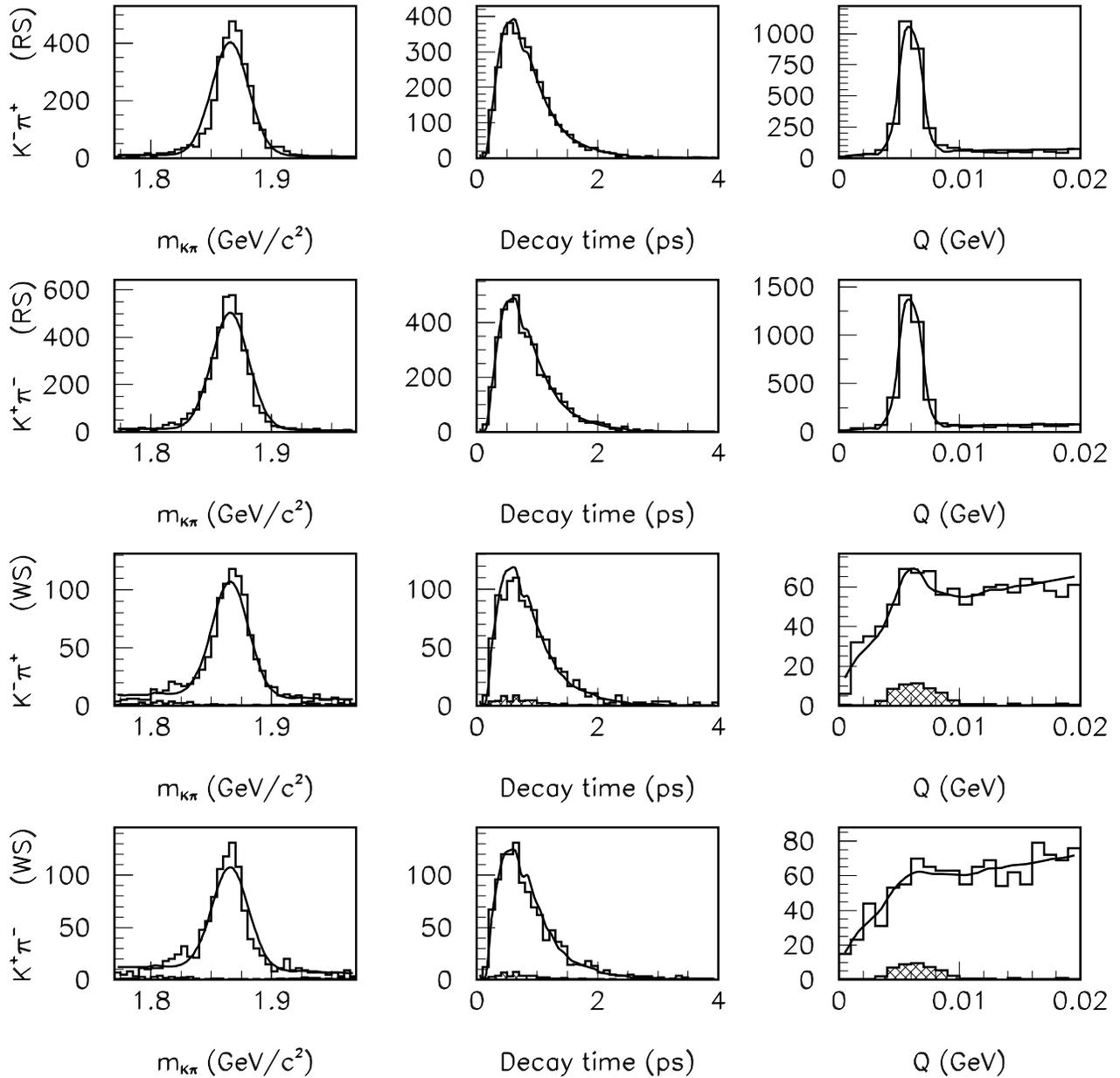,height=17cm,angle=0}}
	\vskip 0.5cm
	\caption{Projections of the four $D\ra K\pi$ data samples
	onto each of three distributions $m_{K\pi}$, $t$ and $Q$.
        Data are from the $D^0$ mass range (1.770 $<$ $m(K\pi)$ 
        $<$ 1.970) GeV/$c^2$. 
	Solid curves show the projections of 
        our primary fit, summarized in 
        Table \protect\ref{tbfitlik}.
	The broad component of the peaks in the WS $Q$ 
        plots is described well by reflections of $K^+K^-$ and
        $\pi^+\pi^-$ signals. MC simulations of reflected $K^+K^-$ and
        $\pi^+\pi^-$ signals normalized to the fit values of
        $A_{KK,\pi\pi}$ are shown as cross-hatched histograms in the 
        wrong-sign plots.
        }
	\label{fgkpioverlay}
\end{figure}

\begin{figure}
	\centering
	\centerline{\psfig{file=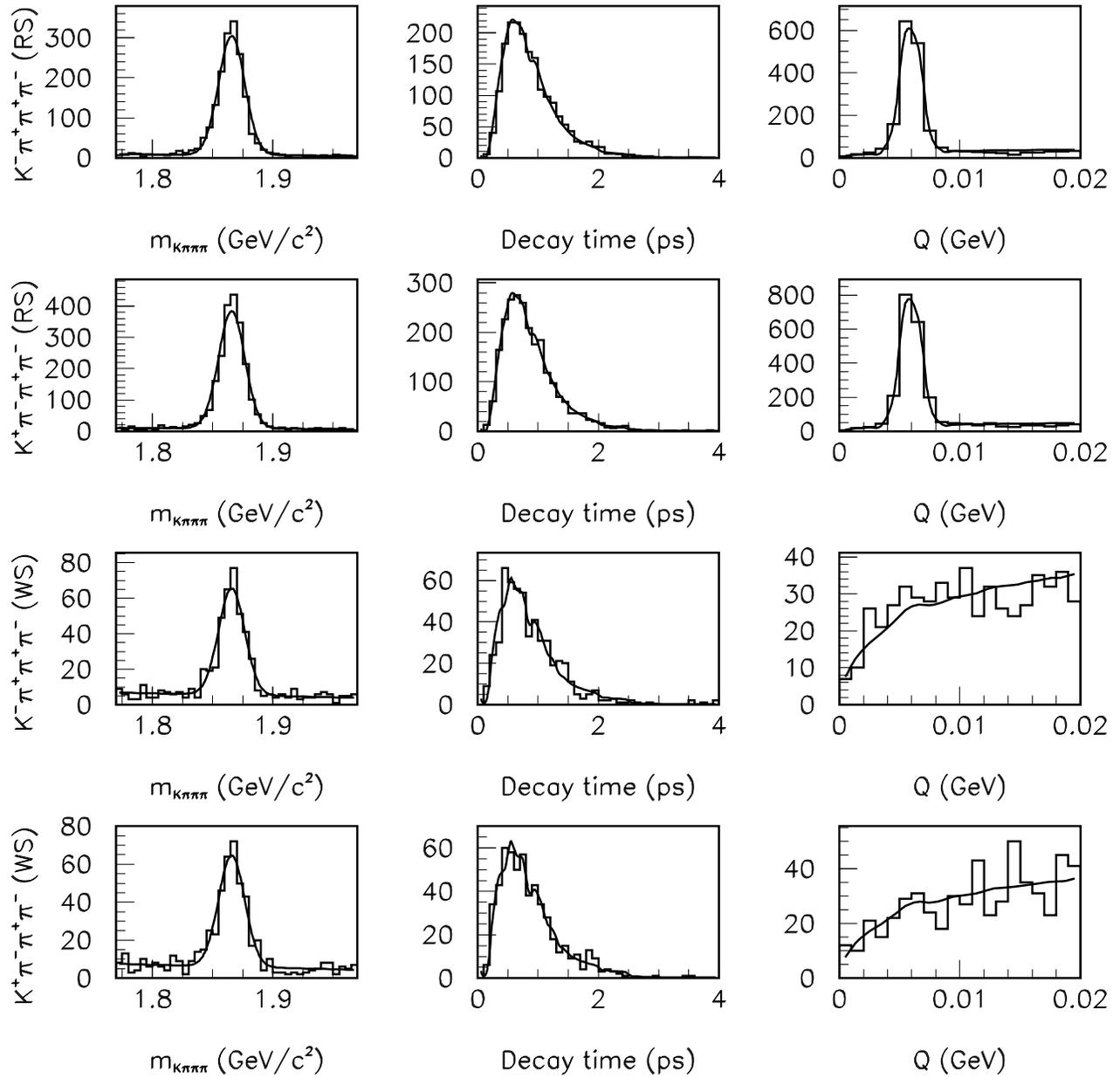,height=17cm,angle=0}}
	\vskip 0.5cm
	\caption{Projections of the four $D\ra K\pi\pi\pi$ data samples
	onto each of three distributions $m_{K\pi\pi\pi}$, $t$, and $Q$.
        Data are from the $D^0$ mass range (1.770 $<$ $m(K\pi\pi\pi)$ 
        $<$ 1.970) GeV/$c^2$. 
	Solid curves show the projections of 
        our primary fit, summarized in 
        Table \protect\ref{tbfitlik}.}
	\label{fgk3pioverlay}
\end{figure}

We fit the data over the range 1.77 to 1.97 GeV/$c^2$ in $m_D$, 0.0 to
0.020 GeV/$c^2$ in $Q$ and 0.0 to 4.0 ps in $t$. Tables~\ref{tbfitlik} and
\ref{tbfitlik_masses} show the resulting 38 parameters from our primary fit,
described in the previous section.
The wrong-sign ratios are all small or
consistent with zero, indicative of small DCS to Cabibbo-favored ratios
and very little mixing.  
Using the criterion $\Delta\ln{{\cal L}}=0.82$ (neglecting systematic errors),
we calculate the one-sided, 90\% C.L. upper limit for mixing to be 
$r_{mix}<0.85$\%.
There is also no evidence for CP violation.
Figures~\ref{fgkpioverlay} and \ref{fgk3pioverlay} show the fit results
overlaid on the data distributions for $m_D$, $t$, and $Q$. Good
agreement is evident in every distribution.

The lego plots of Figure~\ref{fglego} demonstrate that the largest
background comes from real $D$ decays combining with random pions to
produce false $D^*$ candidates. This phenomenon is also apparent in
Figures~\ref{fgkpioverlay} and ~\ref{fgk3pioverlay} where we see many
wrong-sign candidates accumulated at the $\dz$ mass (left column) but
very few of these candidates show the correct Q value (right column) to
have come from $D^*$ decays. A true wrong-sign signal from mixing or DCS
decays would be manifest as a simultaneous peak in both the $M_{K\pi}$
(or $M_{K\pi\pi\pi}$) and Q distributions.

It is important to note that the excess of candidates at Q $\approx$ 0.006
GeV in the wrong-sign decays $\dz\rightarrow K^+\pi^-$ and
$\dzbar\rightarrow K^-\pi^+$ (lower right plots of Figure~\ref{fgkpioverlay}) is due
primarily to $D^*\rightarrow\dz\pi$ with $\dz\rightarrow K^+K^-$ or
$\pi^+\pi^-$, which is misidentified as $\dz\rightarrow K\pi$.  These
candidates are reconstructed at the right Q value for $D^*$ decays, but
appear outside the $\dz$ mass region in $M_{K\pi}$. Although it is hard
to see these candidates in the lego plots of Figure~\ref{fglego}, they
show up as the enhancements in the projected Q distributions in 
Figure~\ref{fgkpioverlay}.

Figure~\ref{fgkpioverlay} also shows the mis-reconstructed $K^+K^-$ and
$\pi^+\pi^-$ mass, time and Q distributions from our Monte Carlo
studies as the cross-hatched histograms in the bottom six plots. The
normalization is determined by our fitted values for $A_{KK,\pi\pi}$ from 
Equation~\ref{eqwsrefl}. Although the reflections are barely visible in
the $m_{K\pi}$ and time distributions, they contribute a broad
enhancement at 0.006 GeV in the Q distributions.
Figure~\ref{fgkkpipi} shows the fitted
contribution of $\dz\rightarrow K^+K^-$ and $\pi^+\pi^-$ misidentified
decays scaled up by a factor of 20 and superimposed on the wrong-sign
mass plot.  The reflected signal is depleted in the $\dz$ mass signal
region (indicated by arrows). More relevant to the mixing rate
determination is Figure~\ref{fgR(Q)}a which shows the combined Q
distributions from $\dz\rightarrow K^+\pi^-$ and $\dzbar\rightarrow
K^-\pi^+$, but with tighter cuts around the $\dz$ mass. Clearly, very
little of the excess remains in the central $\dz$ mass region. The fit
attributes only about 34 candidates to the total wrong-sign $K\pi$ signal.

We have also investigated the effects of other charm backgrounds which
might feed into our wrong-sign samples using Monte Carlo studies and
re-plotting correctly identified states as if they were misidentified.
The largest such source of background comes from doubly-misidentified
decays of $\dz\ra K\pi$ or $K\pi\pi\pi$, in which the $K$ and a $\pi$ of
opposite charge are both misidentified (as $\pi$ and $K$) by the
\cerenkov\ detector.  Although we explicitly cut against these
misidentified decays in our data selection, a small fraction is expected
to pass our cuts.  Our \cerenkov measurements allow $\dz$ decays to be
doubly misidentified around 1.3\% of the time.  However, only about 15\%
of these candidates have an invariant mass within 20 MeV/$c^2$ of the
$\dz$ mass.  The selection cut on the reflected mass of each candidate
further reduces this background by about a factor of 20. Furthermore,
since the background is very broadly distributed in mass, rather than
peaked in the signal region, we expect the fit to respond only weakly by
changing the wrong-sign ratios, probably at the level of a few times
$10^{-4}$ or lower.  Since this background has an exponential decay time
distribution, it will be interpreted as a signal for $r_{dcs}$, and will
not affect the measurement of $r_{mix}$ or $r_{int}$ at all.

The remaining potential sources of charm background come from
$D^*\rightarrow \pi D$ decays with the $D$ decaying to a mode other than
$K\pi$ or $K\pi\pi\pi$. For the $K\pi$ mode, the $\dz\rightarrow K^+K^-$
and $\pi^+\pi^-$ were the most significant, and were handled as
described previously. In addition, we have examined decays
$\dz\rightarrow K^-\pi^+\pi^0$, $K^-\mu^+\nu$ (doubly-misidentified),
and $\dz\rightarrow \pi^+\pi^-\pi^0$ (singly-misidentified), which might
contribute as background to $\dz\rightarrow K\pi$.  As a general rule,
the misidentification rates for these modes are similar to what was
observed for the double misidentification above (all misidentification
is dominated by the \cerenkov selection criteria for the kaon
candidate), while the misidentified masses (in some cases after losing a
neutral particle) are well outside the signal region.  None of the other
decays were seen to contribute a significant background to our data
samples.

\begin{figure}
	\centering
	\centerline{\psfig{file=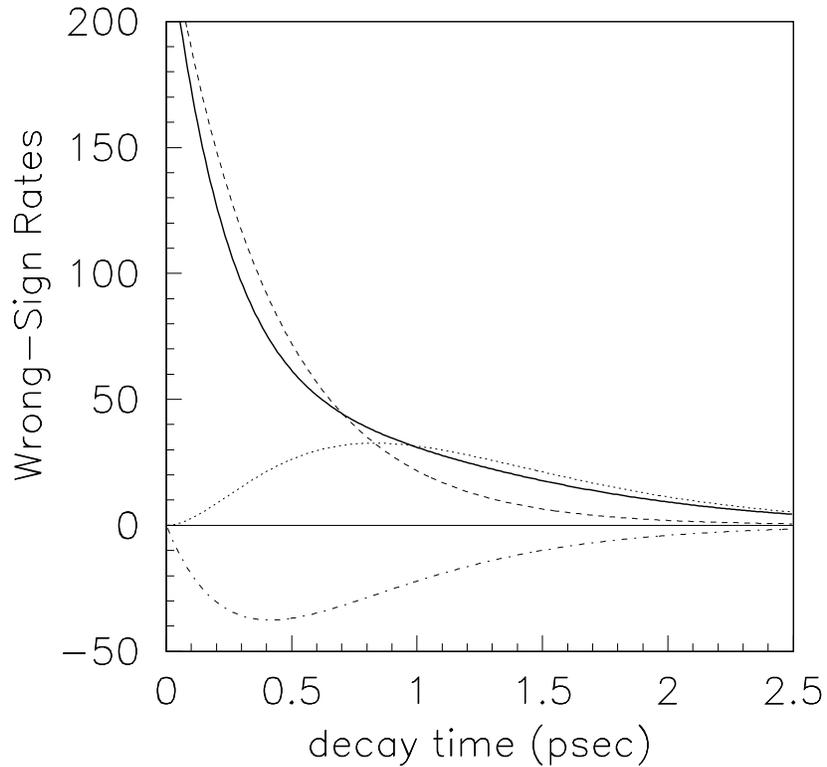,height=10cm,angle=0}}
	\vskip 0.5cm
	\caption{A hypothetical
	plot of the time dependence of wrong-sign decays
        taken from \protect\cite{Blaylock}.
	The dashed line represents the DCS contribution.
	The dotted line shows the contribution due to mixing.
	The dash-dot line shows the
	contribution from destructive interference of DCS and mixing amplitudes
	when the interference is 30\%
	of its maximum.
	The solid line is the sum of all three contributions.
	The vertical scale is arbitrary.}
	\label{fgmixtime}
\end{figure}

In performing the fit, we discovered that the $r_{dcs}$
and $r_{mix}$ terms are strongly anticorrelated with the
$r_{int}$ terms, and strongly correlated with each other.
Figure~\ref{fgmixtime}
demonstrates how these correlations come about
in a hypothetical case where the interference contribution
approximately
cancels the contribution from pure mixing. This plot demonstrates that
even when the full time evolution deviates only slightly from
the pure exponential form of DCS decays, a large contribution
from mixing
can be present if it is offset by a destructive interference
contribution.
This implies that the
fitted values for the
interference contribution and the mixing contribution
are strongly anticorrelated.

\begin{figure}
	\centering
	\vskip 3in
	\centerline{\psfig{file=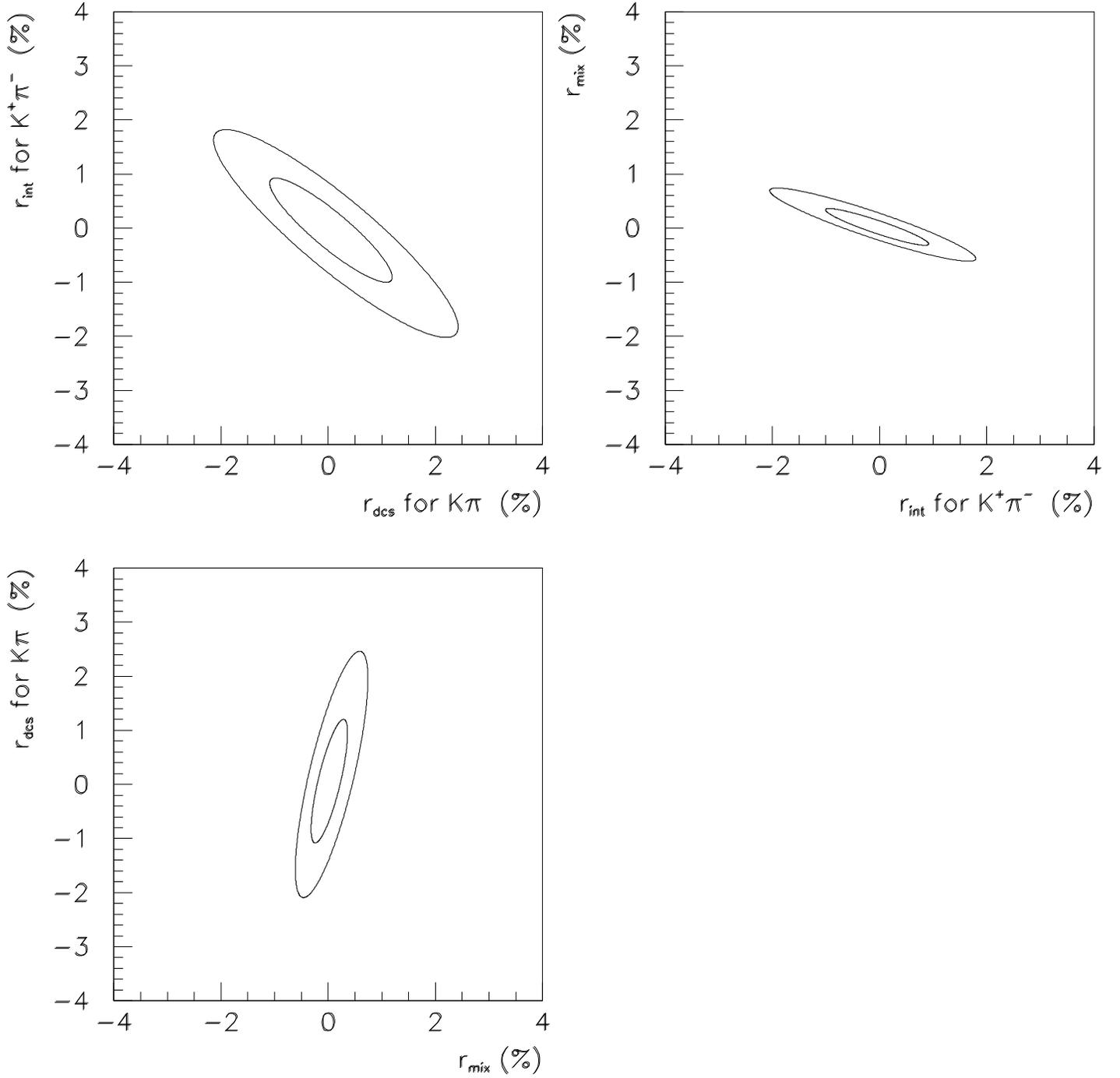,height=19cm,angle=0}}
	\vskip 0.5cm
	\caption{Likelihood contours corresponding to 
	$\Delta\ln{{\cal L}}=0.5$ and $2.0$
	for the fit of Table~\protect\ref{tbfitlik}, 
	illustrating the correlations
	among the three parameters
	$r_{dcs}(K\pi)$, $r_{mix}$ and
	$r_{int}(K^+\pi^-)$. 
	Strong correlations among
	these parameters are apparent. 
	The correlations among
	other wrong-sign ratios are similar.}
	\label{fgcontour}
\end{figure}

Figure~\ref{fgcontour} 
illustrates the correlations in our particular
fit by showing the likelihood contour
plots for representative pairs of parameters for the 
$\dz\ra\kpl\pimi$ mode. 
These strong correlations 
account for much of the uncertainty in the wrong-sign ratios.
Table~\ref{tbcorr} gives the correlation coefficients for the different
wrong-sign ratios. The correlations of these
ratios with all other parameters
of the fit are negligible. We note that the correlations
would be slightly reduced in an experiment with better efficiency
at short decay times where there is good discrimination between
$r_{dcs}$ and the other terms.

\section{Other Fits}\label{sect_other}

Table \ref{tbfitlik} shows our primary results in the search for mixing.
These results assume that CP violation can only occur
in the interference terms of the fit,
an assumption supported by most
extensions to the Standard Model
(see the discussion in Section \ref{sect_CP}). 
However, to answer any concerns about
this assumption, we have also performed a fit in which the CP constraints
are relaxed. Table \ref{tbfitgen} shows the results 
for the wrong-sign ratios of that 41 parameter fit.
As expected, the central values for $r_{dcs}$ and $r_{mix}$
bracket the
corresponding combined terms in Table~\ref{tbfitlik}, and all the
fit errors have increased.
Using the criterion $\Delta\ln{{\cal L}}=0.82$ (neglecting systematic errors),
we calculate the one-sided, 90\% C.L. upper limits to be 
$r_{mix}(\dzbar\ra\dz)<\genlimrzmixzda\%$ and 
$r_{mix}(\dz\ra\dzbar)<\genlimrzmixzdabr\%$.

We note that the earlier measurement by the E691 collaboration
\cite{E691} assumed that the interference terms $r_{int}$
were negligible. Recently, there has been lively discussion concerning
the validity of this assumption \cite{Blaylock,Wolfenstein,Browder}. Although
some authors suggest that the phase between DCS and mixing amplitudes
may be small, and therefore that the interference terms $r_{int}$
should also be small,
we prefer to quote our results without this constraint.
Nonetheless, to compare our measurements with the previous
results from E691, we have performed a fit in which 
we set the interference terms
to zero. Our results for mixing are $r_{mix} = \ncirzmixzda\%$,
which is to be compared with the E691 result 
$r_{mix} = (0.05\pm 0.20)\%$.
The reduction of the fit errors from Table~\ref{tbfitlik}
is indicative of the strong correlations with the $r_{int}$
parameters, which are now fixed at zero. E691  also
touched on this point
by considering several different fixed values
of the interference term (only one interference term was 
allowed in their model).
Their results showed behavior similar
to what we see in our data: strong correlation between 
wrong-sign ratios, and reduced fit errors when fixing the interference
term to zero.
 
\begin{figure}
	\centering
	\vskip 3in
	\centerline{\psfig{file=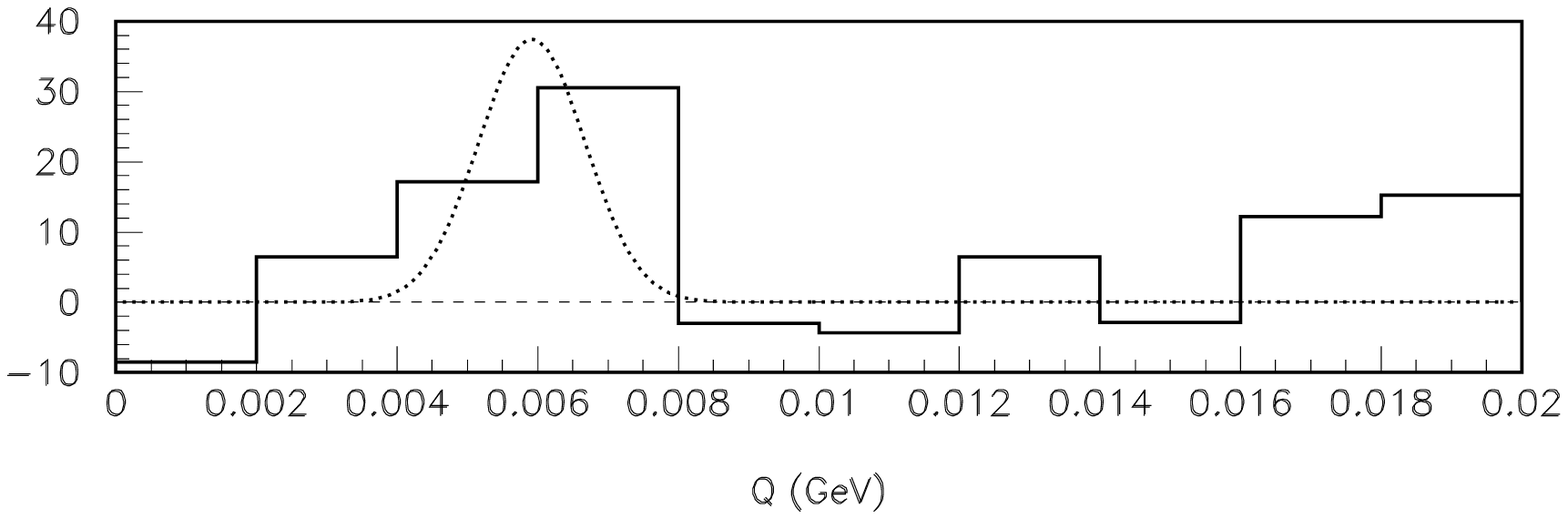,height=3.0in,width=6in,angle=0}}
	\vskip 0.5cm
	\caption{Distribution in Q for wrong-sign $D\ra K\pi$ 
        candidates in the mass range 1.845 to 1.885 GeV/$c^2$.
        The plot has been background subtracted using the fit
        with no mixing (Standard Model case). The Gaussian
        overlay shows the size of the signal attributed to DCS
        decays by the fit.}
	\label{fgbkgsub}
\end{figure}

Finally, we explore the possibility that mixing is completely
negligible, as one would expect from purely Standard Model
contributions. In this case, we fit only for the DCS terms,
obtaining $r_{dcs}(K\pi)=(\nomrzdcskbkm)\%$ and 
    $r_{dcs}(K\pi\pi\pi)=(\nomrzdcskdkm)\%$. 
The result for $\dz\ra K\pi$
demonstrates a two-sigma excess in the signal region which 
we believe is the result of real DCS decays.
Figure \ref{fgbkgsub} shows this excess after background subtraction.
We note that our value for the DCS rate 
is consistent with the CLEO measurement \cite{CLEO} for the total 
wrong-sign rate:
$r_{ws}(K\pi)=(0.77\pm0.25\pm0.25)\%$.

\section{Systematic Uncertainties}\label{sect_syst}

Systematic uncertainties in the fit arise primarily
from our modeling
of the three functions $\epsilon(t)$,
$R(Q)$, and $B_{false}(t)$. By using the right-sign data samples
to estimate these functions, we have minimized our dependence
on Monte Carlo models, but some uncertainty remains.
The results of our studies of systematic uncertainties
are summarized in Tables \ref{tsyslik} and \ref{tsysnom}.
Below we describe the entries in Table~\ref{tsyslik}
which are the systematic errors obtained from 
studies on the fit of Table~\ref{tbfitlik}. The entries in 
Tables \ref{tsysnom} are obtained
similarly. We find
that the systematic uncertainties
in the analysis are small compared with the statistical errors
from the fit.

The uncertainties in the first row arise from our estimates of the
size of the reflection in $K\pi$ from misidentified 
$D^0\to K^-K^+$ and $D^0\to \pi^-\pi^+$ events. Most of this uncertainty
arises from our knowledge of the branching ratios for these modes. We
have estimated this error as 20\%\ of the difference between the results
of fits with and without terms describing these reflections.

\begin{figure}
	\centering
	\centerline{\psfig{file=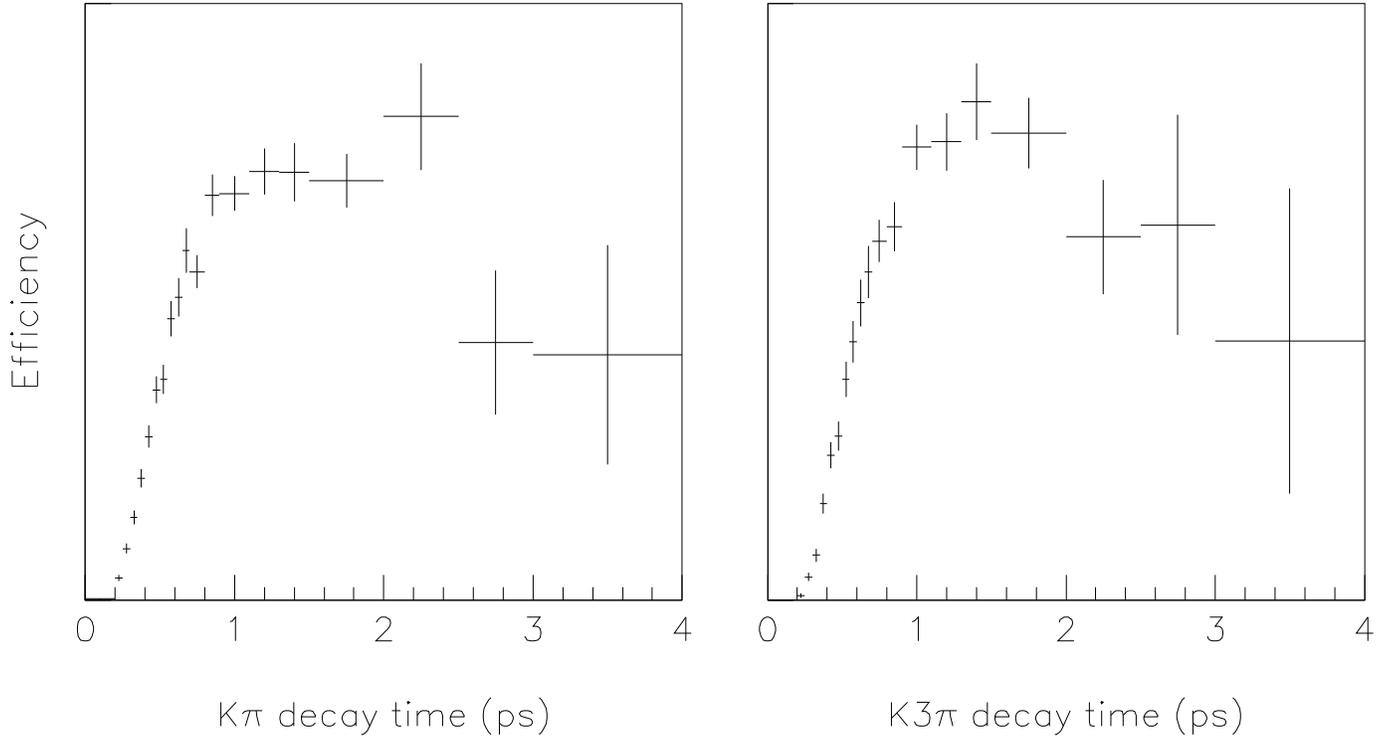,height=10cm,angle=0}}
	\vskip 0.5cm
	\caption{Reconstruction efficiencies for $\dz\ra K\pi$ (left)
		and $\dz\ra K\pi\pi\pi$ (right), as measured from the right-sign
		data samples. The low efficiency at short decay times
		is typical of fixed-target experiments which 
		identify charm decays by a secondary decay vertex.
		The drop in efficiency at long decay times is due
		to our selection criteria which remove decays occuring in
                downstream target foils.
		Efficiencies for charge conjugate final states
		(e.g., $\epsilon_{\kmi\pipl}$ and $\epsilon_{\kpl\pimi}$)
		are observed to be the same within errors, and have
		been combined in the above plots. The vertical scales
		are arbitrary.}
	\label{fgefficiencies}
\end{figure}

\begin{figure}
	\centering
	\vskip 3in
	\centerline{\psfig{file=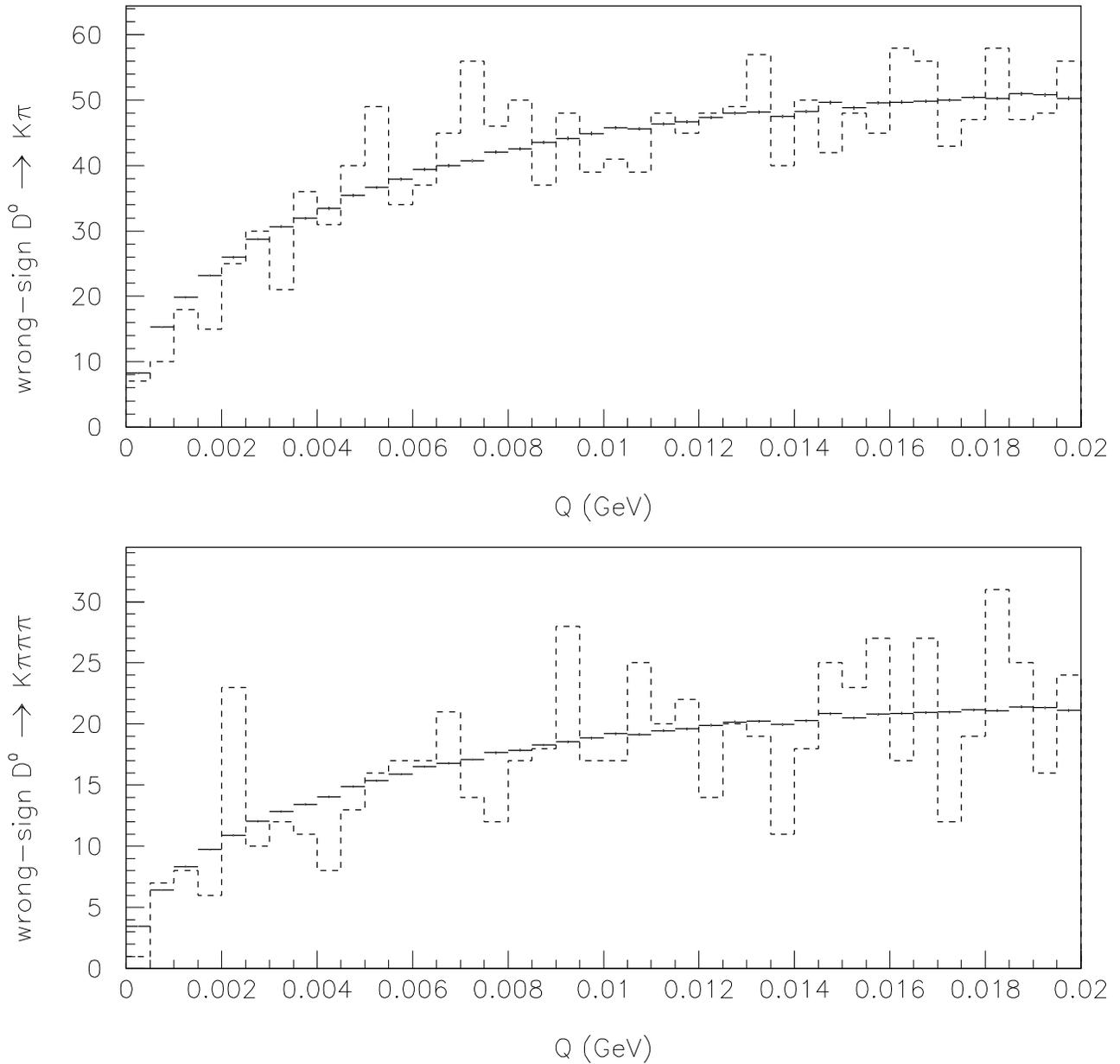,height=17cm,angle=0}}
	\vskip 0.5cm
	\caption{Histograms show distributions of
	$Q\equiv m(K\pi\pi)-m(K\pi)-m(\pi)$ (top) and 
	$Q\equiv m(K\pi\pi\pi\pi)-m(K\pi\pi\pi)-m(\pi)$ (bottom) 
	for wrong-sign $D$ candidates in the mass range
        1.835 to 1.895 GeV/$c^2$.
	The points with error bars show the 
	distributions from 
	combining $\dz$ candidates and $\pi$'s from separate events, normalized
	to the histograms.
	These distributions are used to
	represent $R(Q)$ in the likelihood fits.}
	\label{fgR(Q)}
\end{figure}

\begin{figure}
	\centering
	\centerline{\psfig{file=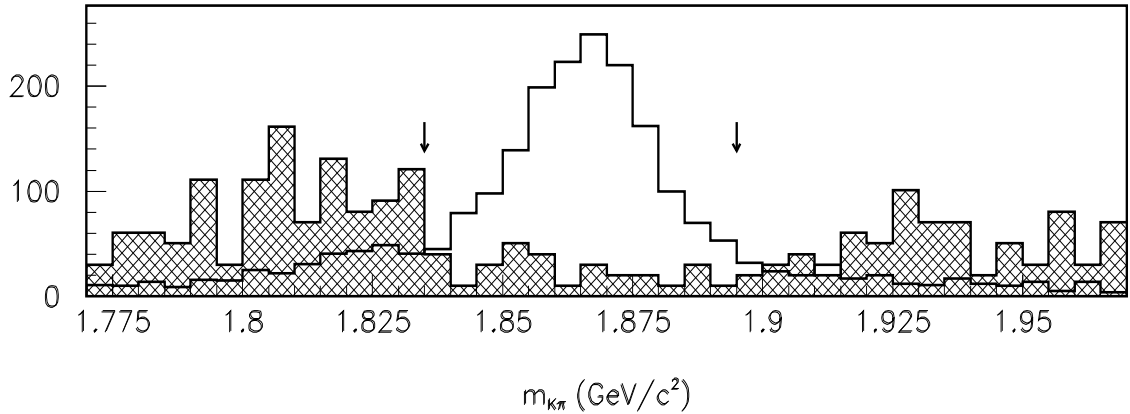,height=2.4in,width=6.0in,angle=0}}
	\vskip 0.5cm
	\caption{The sum of the wrong-sign mass distributions in the
        lower two rows of Figure \protect\ref{fgkpioverlay}. The 
        cross-hatched histogram is the reflected mass distribution of
        $\dz\rightarrow K^+K^-$ and $\dz\rightarrow\pi^+\pi^-$ decays
        from Monte Carlo normalized to 20 times the amount favored by
        our primary fit, summarized in Table \protect\ref{tbfitlik}.
        Notice the depletion of reflected signal
        in the $\dz$ mass signal region. Figure~\protect\ref{fgR(Q)}a 
        shows the Q distribution for events within the restricted mass
        region indicated by the arrows.}
	\label{fgkkpipi}
\end{figure}

The uncertainties listed in the third row 
(``statistics of effy \&\ bkgd distrs'') result from the uncertainties in
our estimates of 
$\epsilon(t)$, $R(Q)$, and $B_{false}(t)$ 
due
to the finite size of our right-sign data sample. The error
bars on the corresponding histograms in Figures~\ref{fgefficiencies},
\ref{fgR(Q)}, and \ref{fgbcomb}  show the
level of uncertainty involved. 
Statistical uncertainties on the
model of the function $R(Q)$ (Figure~\ref{fgR(Q)}) 
have been greatly reduced
by combining each $\dz$ candidate with pions from many different 
events, so that the uncertainties in this histogram are 
negligible in comparison with
the uncertainties of Figures~\ref{fgefficiencies} and \ref{fgbcomb}.
In order to propagate these
statistical uncertainties to uncertainties on the fitted parameters,
we perform many fits to the data, modifying each bin in each of 
the histograms of $\epsilon(t)$, $R(Q)$ and $B_{false}(t)$  
by a Gaussian fluctuation with 
resolution given by the error bars. 
The rms spreads of 
the fitted parameters from
25 such fits are
given in Table~\ref{tsyslik}.
 
The uncertainties listed in the fourth row of Table~\ref{tsyslik}
(``binning of effy \&\ bkgd distrs'') are due to the fact that we
have represented the $\epsilon(t)$, $R(Q)$, and $B_{false}(t)$ functions
by binned histograms rather than smooth functions.  We have, of course,
tried to choose bin sizes small enough so that binning effects are not
significant. In order to verify this claim, we replaced the histograms
in the fit with smoothed functions which were derived from the histogram
data, and repeated the fit. The differences between the parameter values
with the smoothed functions and the parameter values with the histograms
are quoted in Table~\ref{tsyslik} as the uncertainties due to binning.
We expect this method to give an overestimate of the binning effect,
since it also includes the effect of some statistical fluctuations in
the measured histograms which are adjusted by the smoothing function.
As anticipated, the binning effects are small.

The uncertainties listed in the fifth
row of the table (``time resolution'') are due to the
resolution on the measured decay time. 
Since the smearing
is small, a good assumption is that $\epsilon(t)$ is almost the same
with and without smearing. 
For this analysis, the most likely resolution on the decay time 
is about 0.03 ps, with some measurements having a resolution as large as
0.08 ps.
In order to quantify the error due to smearing, we replace the functions of
Equation~(\ref{Bdef}) by functions convoluted with a fixed Gaussian resolution:
	\begin{eqnarray}
	\label{Bdefres}
	B_{exp}(t_i) &=& B_{exp}^0\ \int dt\ e^{-(t-t_i)^2/ 2\sigma_0^2}\
		e^{-\Gamma t}\ \epsilon(t),\nonumber\\
	B_{mix}(t_i) &=& B_{mix}^0\ \int dt\ e^{-(t-t_i)^2/2\sigma_0^2}\
		t^2\ e^{-\Gamma t}\ \epsilon(t),\\
	B_{int}(t_i) &=& B_{int}^0\ \int dt\ e^{-(t-t_i)^2/2\sigma_0^2}\
		t\ e^{-\Gamma t}\ \epsilon(t),\nonumber
	\end{eqnarray}
where $\epsilon(t)$ is obtained as in Section V and the integrals are
obtained numerically.
We then perform the fit with three different values for 
$\sigma_0$: 0.02 ps, 0.05 ps and 0.08 ps. We quote the average of the
central value differences (from the fits for 0.02 ps and 0.05 ps and
from the fits for 0.08 ps and 0.05 ps) as 
variations in Table~\ref{tsyslik}, line 5. 
When the exponential lifetime is modified by our detector acceptance
which is poor at low lifetimes, there is a ``peak'' at $\approx 0.5$ ps.
We observe that the time smearing affects the 
likelihood most near this ``peak'', while mixed events are 
most likely around higher values of decay time.
Consequently, the DCS ratios 
exhibit the largest variation, 
while the mixing ratio is relatively stable.

The uncertainties in the sixth line of the table (``mass resolution'')
come from the assumption of a constant Gaussian resolution in
$m(K\pi)$ and $m(K\pi\pi\pi)$.
In truth, the mass resolution should depend on the $\dz$ momentum
and on the kinematics of the decay. We have studied the dependence
on momentum and verified a noticable correlation between resolution
and momentum. For the $K\pi$ decay mode, about 90\% of the events
have a mass resolution between 12 and 16 MeV/$c^2$, with a 
tail reaching out to about 25 MeV/$c^2$ at high momentum. 
For the $K\pi\pi\pi$ mode, the variation is much smaller, with all events
exhibiting a mass resolution in the range 8.5 to 11 MeV/$c^2$.
To quantify this effect, we have varied the mass resolutions
$\pm 2$ MeV/$c^2$ in the fit and recorded the maximum variations
in wrong-sign ratios in Table~\ref{tsyslik}. The fit results
are quite insensitive to variation of the resolution on $m(K\pi\pi\pi)$,
but change slightly with the resolution on $m(K\pi)$.
A change in $m(K\pi)$ resolution primarily affects $r_{dcs}(K\pi)$, but
because of the correlations in the fit, it will also alter
$r_{mix}$ and $r_{dcs}(K\pi\pi\pi)$, as shown.

Some other assumptions and biases in our fit model
bear further comment. First of all, we have assumed that
the efficiency function $\epsilon(t)$ for reconstructing a $\dz$
from a $\dstar$ decay (signal terms from Eq.~(\ref{eqwssig})) is the same
as the efficiency function for reconstructing a primary $\dz$
(random pion term from Eq.~(\ref{eqwsranpi})). Since our reconstruction
and selection criteria are only weakly dependent on the $\dz$ production 
kinematics, we expect to find very little difference in efficiency
for these two sources of $D$ mesons. Studies of reconstructed $\dz$
decays which are {\it not} associated with a bachelor pion appear to 
confirm that the difference is negligible. Secondly, the $K\pi\pi\pi$
final state may result from different resonant substructures in the
Cabibbo-favored and DCS amplitudes. This can, in principle, lead to
different efficiencies for the DCS and interference terms in the fit.
(The two-body $K\pi$ mode is, of course, immune to this problem.)
Once again however, the fact that our reconstruction depends only
slightly on the decay kinematics leads to effects 
at the level of only 1\%. 
The very similar time dependence of the efficiency functions for the $K\pi$
and $K\pi\pi\pi$ final states (Fig.~\ref{fgefficiencies}) 
demonstrates how little $\epsilon(t)$
depends on the $D$ decay. 

      We are also aware that training the $\dstar$
neural net on a sample of right-sign $\dstar$ decays can,
in principle, produce a small bias in that sample (but not the
wrong-sign samples which were not used for training the net). Careful selection
of input variables for the neural net that are not correlated with
$Q$ (our variables depend only on parameters that describe the bachelor pion)
should prevent any significant bias. 
We have investigated 
this effect by subdividing the training sample into 10 subsamples,
training a neural net on one subsample, and applying
the resulting net to the remaining samples as a test of the bias.
We then repeat the process on each subsample to get a 
better statistical average.
By comparing the sensitivity (measured as $S/\sqrt{B}$) of the 
training samples with the sensitivity of the test samples,
we determine the level of bias.
We find that the number of right-sign signal events
could be biased upwards by about 1\%. This is a negligible
effect compared to our statistical error.

The last line in Table~\ref{tsyslik} shows the contribution from all
the systematic errors added in quadrature. These totals
are less than half the size of the statistical errors in Table~\ref{tbfitlik}.

\section{Discussion}
 
At the current level of sensitivity, mixing searches 
begin to constrain some models\cite{Babu}.
There are also other search methods that are promising.
Using the same $\dstar$ decay chain to identify the produced $D$ meson,
but looking at semileptonic decays of the $D$, is one possibility. 
Although semileptonic decays are harder to reconstruct due to
missing neutrinos, they are not subject to contributions from
DCS amplitudes, and therefore do not suffer from the main limitations
discussed in this paper. 
In a separate publication we describe such a search 
\cite{semi791} with the result $r_{mix}<0.50\%$ at the 90\%\ CL. 
The possibility exists for even higher statistics searches
in future experiments.
Alternatively, it may be possible to detect mixing via the lifetime difference
between the two physical eigenstates by comparing
the measured lifetimes for different CP final states \cite{Liu}. 
Of course, this approach
will only detect mixing if it is associated with a substantial 
lifetime difference
as opposed to mixing that only results from a mass difference.
We are investigating this method as well.
Finally, the cleanest signal for mixing might be found at
a $\tau$-charm factory which produces \d0d0bar\ pairs on resonance.
As has been discussed previously \cite{Bose-stats}, certain hadronic
final states from these \d0d0bar\ pairs can only be produced by mixing
and not by DCS amplitudes. We remain hopeful that one of these techniques
may be used to detect \dmix, and thus provide information about the
existence of new physics.

\section{Summary}
 
      We have searched for evidence of \dmix\ and DCS decays 
by looking for wrong-sign decays in the decay chain
$\dstar\ra\pi D$ with $D\ra K\pi$ or $D\ra K\pi\pi\pi$. Our results are
summarized in Table~\ref{tbsumm}.

      We have seen no
evidence for mixing in either $D^0$ decay mode. The results of a maximum 
likelihood fit to the data are given in Table~\ref{tbfitlik}. The possibility 
of additional sources of wrong-sign decays from DCS amplitudes 
limits our sensitivity for detecting mixing alone.
Using the criterion $\Delta\ln{{\cal L}}=0.82$, 
we calculate the one-sided, 90\% C.L. upper limit to be 
$r_{mix}<\liklimrzmixzda\%$.
If, in order to account for the most general case possible,
we relax the assumption that CP is conserved
in the mixing and DCS terms of the fit (as in Table~\ref{tbfitgen}),
we calculate the upper limits for mixing to be
$r_{mix}(\dzbar\ra\dz)<\genlimrzmixzda\%$ and 
$r_{mix}(\dz\ra\dzbar)<\genlimrzmixzdabr\%$. 

      Our quoted sensitivity to mixing is similar to that of 
Fermilab E691, but our analysis is notably more general in its assumptions 
concerning DCS-mixing interference and CP violation. Assuming no 
DCS-mixing
interference constrains the mixing and DCS contributions much more 
severely, but we do not feel this assumption is justifiable.
Nevertheless, for comparison we include the mixing results for this case
in Table \ref{tbsumm}.
All our results for the $K\pi$ final state are also consistent with the CLEO 
measurement of $r_{ws}(K\pi)=(0.77\pm0.25\pm0.25)\%$
\cite{CLEO} for wrong-sign decays. In particular, if the
mixing amplitude is set to zero, we find a two-sigma enhancement in the $K\pi$
mode and no significant enhancement in the $K\pi\pi\pi$ mode:
$r_{dcs}(K\pi)=(\nomrzdcskbkm)\%$
and $r_{dcs}(K\pi\pi\pi)=(\nomrzdcskdkm)\%$.

      We gratefully acknowledge the assistance of the staff of Fermilab
and of all the participating institutions. This research was supported
by the Brazilian Conselho Nacional de Desenvolvimento Cient\'ifico e
Technol\'ogico, CONACyT (Mexico), the Israeli Academy of Sciences and
Humanities, the U.S. Department of Energy, the U.S.-Israel Binational
Science Foundation, and the U.S. National Science Foundation. Fermilab
is operated by the Universities Research Association, Inc., under
contract with the United States Department of Energy.


\begin{table}
	\caption{Correction factors for the difference
	in integrated reconstruction efficiencies associated with
	the decay time distributions of the mixing and interference
	terms. See Eq.~(\protect\ref{eqeffcorr}) in the text.}
	\label{tbeffcorr}
	\vskip 0.5cm
	\vskip 3in
	\begin{tabular}{cc}
$c_{mix}(K\pi)$	&\gencmixkbpi\rule[-1pt]{0pt}{13pt}\\
$c_{int}(K\pi)$	&\gencintkbpi\rule[-1pt]{0pt}{13pt}\\
$c_{mix}(K\pi\pi\pi)$&\gencmixkdpi\rule[-1pt]{0pt}{13pt}\\
$c_{int}(K\pi\pi\pi)$&\gencintkdpi\rule[-1pt]{0pt}{13pt}\\
\end{tabular}

\end{table}

\begin{table}
	\caption{Signal and background parameters for the fit
        described in Section \protect\ref{sect_technique}, which
	assumes no CP violation in either the DCS or mixing
	terms. Thus, we have used the constraints
	$r_{dcs}(\dz\ra\kpl\pimi)=r_{dcs}(\dzbar\ra\kmi\pipl)$,
	$r_{dcs}(\dz\ra\kpl\pi\pi\pi)=r_{dcs}(\dzbar\ra\kmi\pi\pi\pi)$,
	and $r_{mix}(\dz\ra\dzbar)=r_{mix}(\dzbar\ra\dz)$.}
	\label{tbfitlik}
	\vskip 0.5cm
	
\begin{tabular}{ccccc} 
  &$ \dz\ra\kmi\pipl $&$ \dzbar\ra\kpl\pimi $&$ \dz\ra\kmi \pi\pi\pi $&$ \dzbar\ra\kpl \pi\pi\pi $\rule{0pt}{13pt}\\[1ex] \hline
$A_{rs}$&$ \likArsddskbkmnos $&$ \likArsddskbkpnos $&$ \likArsddskdkmnos $&$ \likArsddskdkpnos $\rule{0pt}{13pt}\\
$A_\pi $&$ \likArsdpikbkmnos $&$ \likArsdpikbkpnos $&$ \likArsdpikdkmnos $&$ \likArsdpikdkpnos $\rule{0pt}{13pt}\\
$A_0 $&$ \likArsbpikbkmnos $&$ \likArsbpikbkpnos $&$ \likArsbpikdkmnos $&$ \likArsbpikdkpnos $\rule{0pt}{13pt}\\[1ex] \hline
  &$ \dzbar\ra\kmi\pipl $&$ \dz\ra\kpl\pimi $&$ \dzbar\ra\kmi \pi\pi\pi $&$ \dz\ra\kpl \pi\pi\pi $\rule{0pt}{13pt}\\[1ex] \hline
$r_{dcs}(\%)$&\multicolumn{2}{c}{$ \likrzdcskbkm $}&\multicolumn{2}{c}{$ \likrzdcskdkm $}\rule{0pt}{13pt}\\
$r_{mix}(\%)$&\multicolumn{4}{c}{$ \likrzmixzda $}\rule{0pt}{13pt}\\
$r_{int}(\%)$&$ \likrzintkbkm $&$ \likrzintkbkp $&$ \likrzintkdkm $&$\likrzintkdkp $\rule{0pt}{13pt}\\
$A_\pi $&$ \likAwsdpikbkmnos $&$ \likAwsdpikbkpnos $&$ \likAwsdpikdkmnos $&$
\likAwsdpikdkpnos $\rule{0pt}{13pt}\\
$A_0 $&$ \likAwsbpikbkmnos $&$ \likAwsbpikbkpnos $&$ \likAwsbpikdkmnos $&$
\likAwsbpikdkpnos $\rule{0pt}{13pt}\\
$A_{KK,\pi\pi} $&$ \likAzrefzkmnos $&$ \likAzrefzkpnos $& \omit &
\omit \rule{0pt}{13pt}\\[1ex] \hline
  &$ \kmi\pipl  $&$ \kpl\pimi $&$ \kmi \pi\pi\pi  $&$ \kpl \pi\pi\pi  $\rule{0pt}{13pt}\\[1ex] \hline
$A_1$ ($c^2$/GeV) &$\likslopezkbkmnos $&$ \likslopezkbkpnos $&$\likslopezkdkmnos
$&$\likslopezkdkpnos $\rule{0pt}{13pt}\\[1ex] 
\end{tabular}

\end{table}

\begin{table}
	\caption{Mass parameters for the fit
        described in Section \protect\ref{sect_technique}, which
        assumes no CP violation in either the DCS or mixing
        terms. Thus, we have used the constraints
        $r_{dcs}(\dz\ra\kpl\pimi)=r_{dcs}(\dzbar\ra\kmi\pipl)$,
        $r_{dcs}(\dz\ra\kpl\pi\pi\pi)=r_{dcs}(\dzbar\ra\kmi\pi\pi\pi)$,
        and $r_{mix}(\dz\ra\dzbar)=r_{mix}(\dzbar\ra\dz)$.
        All values are in
	MeV/$c^2$. There are systematic uncertainties on these
	parameters that are bigger than the statistical errors shown here,
	but they have inconsequential effects on the parameters 
	of Table~\protect\ref{tbfitlik}.}
	\label{tbfitlik_masses}
	\vskip 0.5cm
	
\begin{tabular}{cc}

$m_\dz$		&$\likdazmassnos$         \rule[-1pt]{0pt}{13pt}\\
$\sigma_{K\pi}$	&$\likKbpizsigmanos$      \rule[-1pt]{0pt}{13pt}\\
$\sigma_{K\pi\pi\pi}$&$\likKdpizsigmanos$ \rule[-1pt]{0pt}{13pt}\\
$Q_\dstar$	&$\likdstarzqanos$        \rule[-1pt]{0pt}{13pt}\\
$\sigma_Q$	&$\likdstarzsigmnos$      \rule[-1pt]{0pt}{13pt}\\

\end{tabular}

\end{table}

\begin{table}
      \caption{Correlation coefficients for the wrong-sign
      ratios from the fit of Table~\protect\ref{tbfitlik}. Only
      the lower halves of the symmetric matrices are shown.
      Correlations with
      the fit parameters that are not shown are negligible.}
      \label{tbcorr}
      \vskip 0.5cm
      \begin{tabular}{cccccccc}
 &$r_{dcs}(K\pi)$ &$r_{dcs}(K\pi\pi\pi)$ &$r_{int}(\kmi\pipl)$ &$r_{int}(\kpl\pimi)$ &$r_{int}(\kmi\pi\pi\pi)$ &$r_{int}(\kpl\pi\pi\pi)$ &$r_{mix}$
\rule{0pt}{13pt}\\[1ex]\hline
$r_{dcs}(K\pi)$ 
&\likcoefdada&&&&&&
\rule[-1pt]{0pt}{13pt}\\
$r_{dcs}(K\pi\pi\pi)$ 
&\likcoefdcda&\likcoefdcdc&&&&&
\rule[-1pt]{0pt}{13pt}\\
$r_{int}(\kmi\pipl)$ 
&\likcoefdeda&\likcoefdedc&\likcoefdede&&&&
\rule[-1pt]{0pt}{13pt}\\
$r_{int}(\kpl\pimi)$ 
&\likcoefdfda&\likcoefdfdc&\likcoefdfde&\likcoefdfdf&&&
\rule[-1pt]{0pt}{13pt}\\
$r_{int}(\kmi\pi\pi\pi)$ 
&\likcoefdgda&\likcoefdgdc&\likcoefdgde&\likcoefdgdf&\likcoefdgdg&&
\rule[-1pt]{0pt}{13pt}\\
$r_{int}(\kpl\pi\pi\pi)$ 
&\likcoefdhda&\likcoefdhdc&\likcoefdhde&\likcoefdhdf&\likcoefdhdg&\likcoefdhdh&
\rule[-1pt]{0pt}{13pt}\\
$r_{mix}$
&\likcoefdida&\likcoefdidc&\likcoefdide&\likcoefdidf&\likcoefdidg&\likcoefdidh&\likcoefdidi
\rule[-1pt]{0pt}{13pt}\\
\end{tabular}

\end{table}

\begin{table}
	\caption{Fit results for the wrong-sign ratios of the
        most general fit, with no assumptions about CP.}
	\label{tbfitgen}
	\vskip 0.5cm
	
\begin{tabular}{ccccc}
 &$ \dzbar\ra\kmi\pipl $&$ \dzbar\ra\kmi\pi\pi\pi $&$ \dz\ra\kpl\pimi $&$ \dz\ra\kpl\pi\pi\pi $\rule{0pt}{13pt}\\[1ex]\hline
$r_{dcs}(\%)$&$ \genrzdcskbkm $&$ \genrzdcskdkm $&$ \genrzdcskbkp $&$
\genrzdcskdkp $\rule{0pt}{13pt}\\
$r_{mix}(\%)$&\multicolumn{2}{c}{$ \genrzmixzda $}
      &\multicolumn{2}{c}{$ \genrzmixzdabr $}\rule{0pt}{13pt}\\
$r_{int}(\%)$&$ \genrzintkbkm $&$ \genrzintkdkm $&$ \genrzintkbkp $&$
\genrzintkdkp $\rule{0pt}{13pt}\\
\end{tabular}

\end{table}

\begin{table}
	\caption{Systematic uncertainties for the key parameters
        in the fit of 
	Table~\protect\ref{tbfitlik}. This fit describes the case
        where we have allowed for CP violation
        only in the interference term.
        Entries are explained in the text.}
	\label{tsyslik}
	\vskip 0.5cm
	\begin{tabular}{cccc}
			&$r_{dcs}(K\pi)\ (\%)		$&$r_{dcs}(K\pi\pi\pi)\ (\%)	$&$r_{mix}\ (\%)		$
\rule{0pt}{13pt}\\[1ex]\hline
fit value		&$\likrzdcskbkmnos$&$\likrzdcskdkmnos$&$\likrzmixzdanos$
\rule[-1pt]{0pt}{13pt}\\
$K^+K^-$, $\pi^+\pi^-$ reflections &$ \pm\liksyskbpref $&$\pm\liksyskdpref $&$ \pm\liksysmixref$
\rule[-1pt]{0pt}{13pt}\\
statistics of effy \&\ bkgd distrs &$ \pm\liksyskbpsta $&$ \pm\liksyskdpsta $&$ \pm\liksysmixsta$
\rule[-1pt]{0pt}{13pt}\\
binning of effy \&\ bkgd distrs    &$ \pm\liksyskbpbin $&$ \pm\liksyskdpbin $&$ \pm\liksysmixbin$
\rule[-1pt]{0pt}{13pt}\\
time resolution	 &$ \pm\liksyskbptim $&$ \pm\liksyskdptim $&$ \pm\liksysmixtim$
\rule[-1pt]{0pt}{13pt}\\
mass resolution	 &$ \pm\liksyskbpmas $&$ \pm\liksyskdpmas $&$ \pm\liksysmixmas$
\rule[-1pt]{0pt}{13pt}\\[1ex]\hline
total systematic uncertainty &$ \pm\liksyskbptot $&$ \pm\liksyskdptot $&$ \pm\liksysmixtot$
\rule[-1pt]{0pt}{13pt}\\

\end{tabular}

\end{table}

\begin{table}
	\caption{Systematic uncertainties in the no-mixing
        fit, corresponding to the Standard Model case. 
	Entries are explained in the text.}
	\label{tsysnom}
	\vskip 0.5cm
	\begin{tabular}{ccc}
			&$r_{dcs}(K\pi)\ (\%)		$&$r_{dcs}(K\pi\pi\pi)\ (\%)	$
\rule{0pt}{13pt}\\[1ex]\hline
fit value		&$\nomrzdcskbkmnos$&$\nomrzdcskdkmnos$
\rule[-1pt]{0pt}{13pt}\\
$K^+K^-$, $\pi^+\pi^-$ reflections &$ \pm\nomsyskbpref $&$\pm\nomsyskdpref $
\rule[-1pt]{0pt}{13pt}\\
statistics of effy \&\ bkgd distrs &$ \pm\nomsyskbpsta $&$ \pm\nomsyskdpsta $
\rule[-1pt]{0pt}{13pt}\\
binning of effy \&\ bkgd distrs    &$ \pm\nomsyskbpbin $&$ \pm\nomsyskdpbin $
\rule[-1pt]{0pt}{13pt}\\
time resolution	 &$ \pm\nomsyskbptim $&$ \pm\nomsyskdptim $
\rule[-1pt]{0pt}{13pt}\\
mass resolution	 &$ \pm\nomsyskbpmas $&$ \pm\nomsyskdpmas $
\rule[-1pt]{0pt}{13pt}\\[1ex]\hline
total systematic uncertainty &$ \pm\nomsyskbptot $&$ \pm\nomsyskdptot $
\rule[-1pt]{0pt}{13pt}\\

\end{tabular}

\end{table}

\begin{table}
	\caption{A summary of values from our four fits. 
	The top line describes the most likely case for extensions
	to the Standard Model which produce large mixing.
	The second line describes our most general fit, providing
	the most conservative results.
	The third line matches the assumptions of previous experiments,
	which we do not feel are justifiable.
	The bottom line describes the Standard Model case.
        Results from other experiments are also listed for comparison.}
	\label{tbsumm}
	\vskip 0.5cm
	\begin{tabular}{ccc}
Fit type      & This Result      & Other Comparable Result 
\rule{0pt}{13pt}\\[1ex]\hline

CP violation only in & $r_{mix} = (\likrzmixzda)\%$ & $r_{mix} =
(0.11^{+0.30}_{-0.27})$\% (E791 \cite{semi791})
\rule[-1pt]{0pt}{7pt}\\
interference term & \omit & (Semileptonic decays)
\rule{0pt}{13pt}\\[1ex]\hline
Most general, & $r_{mix}(\dzbar\ra\dz) = (\genrzmixzda)\%$ & \omit
\rule[-1pt]{0pt}{7pt}\\
no CP assumptions & $r_{mix}(\dz\ra\dzbar) = (\genrzmixzdabr)\%$ & \omit
\rule{0pt}{13pt}\\[1ex]\hline
No CP violation, & $r_{mix} = (\ncirzmixzda)\%$ & $r_{mix} = (0.05
\pm 0.20)\%$ (E691 \cite{E691})
\rule[-1pt]{0pt}{7pt}\\
no interference  & \omit              & \omit 
\rule{0pt}{13pt}\\[1ex]\hline
No mixing        & $r_{dcs}(K\pi)=(\nomrzdcskbkm)\%$ &
                                    $r_{ws}(K\pi)=(0.77\pm0.25\pm0.25)\%$ 
\rule[-1pt]{0pt}{7pt}\\
\omit            & $r_{dcs}(K\pi\pi\pi)=(\nomrzdcskdkm)\%$ & (CLEO \cite{CLEO})
\rule[-1pt]{0pt}{13pt}\\

\end{tabular}

\end{table}

%
%

%
%

%
%

\vskip 9.0cm

\begin{references}
\bibitem{CFCPinv}
We have assumed in this expression that the Cabibbo-favored
amplitudes, which are described by Standard Model processes,
are dominated by a single weak phase. Thus,
$\bra{f}H\ket{\dzbar}_{CF}=\bra{\bar f}H\ket{\dz}_{CF}$ to
good approximation. This will also be of relevance in our
interpretation of 
$r_{mix}$, discussed later in this paper.
\bibitem{tsmear}
We have made the simplifying assumption that the efficiency 
can be expressed as a function of $t$ alone. In truth, it
can depend on a number of parameters involving the location
of the decay vertex, the $D$ momentum, and the decay kinematics.
Consequently, our formalism is rigorously correct only for the DCS
term, which has the same time distribution as the right-sign sample
used to measure $\epsilon(t)$. However, in this analysis,
we believe the efficiency is mostly dependent on the
decay time of the $\dz$ and is therefore the same for
all three time distributions of Eq.~(\ref{Bdef}).  Of course, the
efficiency may more correctly be described as a function of the true
decay time instead and should therefore be included inside the integrals
of Eq.~(\ref{Bdef}). Since the smearing is small, we do not do so and
discuss the associated systematic error in Section VII below.
\bibitem{weightedsum}
Instead of constraining $r_{mix}$ to be the same for both $K\pi$
and $K\pi\pi\pi$ modes, we could of course fit the modes separately
and combine the results to obtain an estimate
of the common value of $r_{mix}$. However, because of the strong
correlations present, it is easier to let the fit
perform the constraint than to propagate the full correlation
matrices in a weighted mean of the separate modes. The results
are essentially the same in either case.
\end{references}
\end{document}